\documentclass[useAMS,twocolumn,usenatbib]{mn2e}
\usepackage{times,psfig,epsfig} 
\usepackage{rotating, graphicx}
\usepackage{color}
\usepackage{amssymb, amsmath}
\usepackage[T1]{fontenc} 
\usepackage{aecompl}

\newcommand{\chandra}{{\it Chandra}}

\newcommand{\rosat}{{\it ROSAT}}
\newcommand{\xmm}{{\it XMM-Newton}}
\newcommand{\rexcess}{{\it REXCESS}}
\newcommand{\planck}{{\it Planck}}
\newcommand{\subaru}{{\it Subaru}}
\newcommand{\sza}{{\it Sunyaev-Zel'dovich Array}}

\newcommand{\nr}{NR}
\newcommand{\w}{CSF}
\newcommand{\agn}{AGN}

\newcommand{\be}{\begin{equation}}
\newcommand{\ee}{\end{equation}}
\newcommand{\ba}{\begin{eqnarray}}
\newcommand{\ea}{\end{eqnarray}}
\newcommand{\brr}{\begin{array}}
\newcommand{\err}{\end{array}}
\newcommand{\bc}{\begin{center}}
\newcommand{\ec}{\end{center}}

\newcommand{\msun}{\,h^{-1}M_\odot}

\newcommand{\yx}{\mbox{$Y_{\rm{X}}$}}
\newcommand{\ysz}{\mbox{$Y_{\rm{SZ}}$}}

\newcommand{\mincir}{\raise
  -2.truept\hbox{\rlap{\hbox{$\sim$}}\raise5.truept \hbox{$<$}\ }}
\newcommand{\magcir}{\raise
  -2.truept\hbox{\rlap{\hbox{$\sim$}}\raise5.truept \hbox{$>$}\ }}
\newcommand{\siml}{\raise
  -2.truept\hbox{\rlap{\hbox{$\sim$}}\raise5.truept \hbox{$<$}\ }}
\newcommand{\simg}{\raise
  -2.truept\hbox{\rlap{\hbox{$\sim$}}\raise5.truept \hbox{$>$}\ }}

\title[Pressure of the hot gas in simulations of galaxy clusters]
{Pressure of the hot gas in simulations of galaxy clusters}

\author[S. Planelles et al.]
{S. Planelles$^{1,2}$\thanks{e-mail: susana.planelles@uv.es}, D. Fabjan$^{3,5}$\thanks{e-mail: dunja.fabjan@fmf.uni-lj.si}, S. Borgani$^{2,3,4}$, G. Murante$^{3}$, E. Rasia$^{3,12}$, V. Biffi$^{2,3}$,  
\newauthor
 N. Truong$^{6}$,  C. Ragone-Figueroa$^{9,3}$, G.~L. Granato$^{3}$, K. Dolag$^{7,8}$, E. Pierpaoli$^{10}$, 
  \newauthor
A. M. Beck$^{7}$, Lisa K. Steinborn$^{7}$, M. Gaspari$^{11}$
  \\~\\
\footnotesize 
$^1$ Departamento de Astronom{\'i}a y Astrof{\'i}sica, Universidad de Valencia, c/ Dr. Moliner, 50, 46100 - Burjassot (Valencia), Spain\\
$^2$ Astronomy Unit, Department of Physics, University of Trieste, via Tiepolo 11, I-34131 Trieste, Italy\\
$^3$ INAF, Osservatorio Astronomico di Trieste, via Tiepolo 11, I-34131 Trieste, Italy\\ 
$^4$ INFN -- National Institute for Nuclear Physics, Via Valerio 2, I-34127 Trieste, Italy\\ 
$^5$ Faculty of Mathematics and Physics, University of Ljubljana, Jadranska 19, 1000 Ljubljana, Slovenia \\
$^6$ Dipartimento di Fisica, Universit{\`a} di Roma Tor Vergata, via della Ricerca Scientifica, I-00133, Roma, Italy\\
$^7$ University Observatory Munich, Scheinerstr. 1, 81679 Munich, Germany\\
$^8$ Max-Planck-Institut f\"ur Astrophysik, Karl-Schwarzschild Strasse 1, 85748 Garching bei M\"{u}nchen, Germany\\
$^9$ Instituto de Astronom\'ia Te\'orica y Experimental (IATE),\\
  Consejo Nacional de Investigaciones Cient\'ificas y T\'ecnicas de la Rep\'ublica Argentina (CONICET),\\ Observatorio
  Astron\'omico, Universidad Nacional de C\'ordoba, Laprida 854, X5000BGR, C\'ordoba, Argentina\\
$^{10}$ University of Southern California, Los Angeles, CA 90089\\
$^{11}$ Department of Astrophysical Sciences, Princeton University, Princeton, NJ 08544, USA; Einstein and Spitzer Fellow\\
$^{12}$ Department of Physics, University of Michigan, 450 Church St., Ann Arbor, MI 48109
 }

\begin{document}
\maketitle 

\begin{abstract}

We analyze the radial pressure profiles, the ICM clumping factor and the Sunyaev-Zel'dovich (SZ) scaling relations of  a sample of simulated  galaxy clusters and groups identified in a set of hydrodynamical simulations based on an updated version of the TreePM--SPH {\footnotesize {\sc GADGET-3}} code.  
Three different sets of simulations are performed: the first assumes non-radiative physics, the others include, among other processes,  
AGN and/or stellar feedback.  
Our results are analyzed as a function of redshift, ICM physics, cluster mass and cluster cool-coreness or dynamical state. 
In general,  the mean pressure profiles obtained for our sample of groups and clusters show a good agreement with  X-ray and SZ observations. 
Simulated cool-core (CC)  and non-cool-core (NCC) clusters also show  a good match with real data. 
We obtain in all cases a small (if any) redshift evolution of the pressure profiles of massive clusters, at least back to z = 1. 
We find that the clumpiness of gas density and pressure increases with the distance from the cluster center and with the dynamical activity. 
The inclusion of AGN feedback in our simulations generates  values for the gas clumping ($\sqrt C_{\rho}\sim 1.2$ at $R_{200}$) in good agreement with recent observational estimates.
The simulated $Y_{SZ}-M$ scaling relations are in good accordance with several observed samples, especially for massive clusters.  
As for the scatter of these relations, we obtain a clear dependence on the cluster dynamical state, whereas this distinction is not so evident when looking at the subsamples of CC and NCC clusters. 
  
\end{abstract} 
 
\begin{keywords}  
cosmology:  methods: numerical -- galaxies: cluster: general -- X-ray: galaxies. 
\end{keywords}

\section{Introduction}
\label{sec:intro}

Galaxy clusters represent ideal systems for studies of both cosmology
and astrophysics \citep[e.g.,][for recent reviews]{Kravtsov_2012,
Planelles_2014}.  Although most of the total cluster mass is in the
form of dark matter (DM; $\sim85\%$), they also contain a
significant baryonic budget formed by the hot intra-cluster medium
(ICM; $\sim12\%$) and the stellar component ($\sim3\%$).  Thanks to
this particular composition, galaxy clusters can be observed in
different wavebands, providing complementary observational probes to
exploit their use as cosmological tools \citep[e.g.][]{Allen2011}.

The ICM, a hot ionized gas with typical temperatures of $10^7-10^8$ K,
emits strongly in the X-ray band via thermal Bremsstrahlung
\citep{Sarazin_1988}. The resulting X-ray surface brightness depends
quadratically on the electron number density $n_e$.  On the other
hand, galaxy clusters can also be observed in the millimeter waveband via
the thermal Sunyaev-Zel'dovich effect \citep[SZ effect;][]{SZ_1972}. In this
case, the observed signal depends on the dimensionless
Compton $y$ parameter, which is proportional to the integral along the
line-of-sight of the electron pressure ($P_e\propto n_eT_e$).
Given their definitions, SZ and X-ray observations of galaxy clusters
are differently sensitive to the details of the gas distribution and
thermal state, thus allowing for 
complementary descriptions of the ICM thermodynamics. To
exploit this connection, the ICM pressure distribution, directly
connected to the depth of the cluster gravitational potential well
and, therefore, to the total cluster mass, is a crucial quantity.   

Both SZ observations and numerical simulations consistently show that
the integrated SZ parameter, \ysz, is tightly related to cluster
mass \citep[][]{Nagai_2006, Bonaldi_2007, Bonamente_2008, Kay_2012,
Battaglia_2012a, Planck_2013_a}. In particular, simulations indicate that,
independently of the cluster dynamical state, the \ysz$-M$ relation
has a low scatter, with its normalization being only slightly
dependent on the ICM physics.  The X-ray analogue of \ysz, defined
as the product of the gas mass and its X-ray temperature,
i.e. $Y_X\equiv M_{gas} \cdot T_X$ \citep{Kravtsov_2006}, has also
been shown to behave in a similar way \citep[e.g.][]{Nagai_2007_2,
Fabjan_2011, Planelles_2013_b, Hahn_2015, Truong_2016}. These two quantities
relate to each other via the cluster thermal pressure distribution.
Therefore, a detailed analysis of the ICM thermal pressure, and its dependencies on
cluster mass, physics and redshift, are crucial to deepening 
our understanding of both ICM physics and cosmology.

The pressure structure of the ICM is  sensitive to the combined
action of gravitational and non-gravitational physical processes
affecting galaxy clusters.
Since these processes, acting on both galactic and cosmological
scales, take place during the cluster evolution, the ICM is hardly ever in
perfect hydrostatic equilibrium \citep[HE; e.g.][]{Rasia_2004, Rasia_2006, Nagai2007, Gaspari_2014, Biffi_2016}.  
The first observational constraints on cluster pressure profiles were provided
by X-ray observations \citep[e.g.,][]{Finoguenov_2007,
Arnaud_2010}. However, given their dependence on the square of the
gas density, only extremely long exposures allow the observations of 
cluster outskirts \citep[][]{Simionescu2011, Urban_2011,
Urban_2014}.  On the contrary, since SZ effect measurements depend linearly
on the gas density and temperature, they are more
suited to probe the external clusters regions \citep[e.g.][]{Plagge_2010,
 Basu_2010, Bonamente_2012, Planck_2013}.  Indeed, thanks to 
SZ facilities, such as the Planck telescope \citep{Planck_2013_c, Bourdin_2015}, the
South Pole Telescope \citep[SPT;][]{Reichardt_2013}, the Atacama
Cosmology Telescope \citep[ACT;][]{ACT_2013}, or the Combined 
Array for Research in Millimeter-wave Astronomy 
\citep[CARMA;][]{Carma_2013}, the number and characterization 
of large SZ-selected cluster surveys has been dramatically improved in the recent past.
From an observational point of view, however, constraining the
outskirts of clusters is still challenging in either wavelengths because 
they require observations with high sensitivity: long exposures to detect low
surface brightness and good angular resolution to remove contributions
from point-like sources \citep[e.g.][]{Reiprich_2013, Planck_Inter_X_2013}.

To date, X-ray and SZ observations of galaxy clusters together with
results from numerical simulations \citep[e.g.,][]{Borgani_2004,
Nagai_2007_2, Piffaretti_2008} suggest that the ICM radial pressure
distribution follows a nearly universal shape
\citep[e.g.][]{Arnaud_2010}, at least out to $R_{500}$ at $z=0$. This
universal pressure profile is well described by a generalized NFW
model {\citep[GNFW;][]{Zhao_1996, Nagai_2007_2}}, which is relatively insensitive
to the details of the ICM physics.  Beyond $R_{500}$ the consistency
of these results still needs to be confirmed since, besides the
observational challenge, a number of processes \citep[such as ongoing
mergers and diffuse accretion, clumpy gas distribution, turbulent
pressure support, e.g.,][]{Lau_2009, Vazza_2009,
Battaglia_2010, Nagai_Lau_2011, Roncarelli_2013, Zhuravleva_2013,
Vazza_2013, Battaglia_2014,Gaspari_2015} contribute to deviate 
clusters from the idealized case of thermal pressure equilibrium in a 
smooth ICM distribution.  As for the redshift evolution of the universal 
pressure profile, simulations predict no significant evolution outside of the
cluster core \citep[e.g.][]{Battaglia_2012a}, a result that is
confirmed by recent observations of high-redshift clusters
\citep[][]{McDonald_2014, Adam_2014}.

In order to explain possible deviations from a universal pressure
profile, a detailed modeling of the gas density and pressure clumping
from the core out to the cluster outskirts is of particular interest.
Gas density and pressure clumpiness impact on X-ray and SZ effect
measurements. For instance, gas inhomogeneities can bias high
measurements of the gas density profiles
\citep[e.g.][]{Mathiesen_1999, Nagai_Lau_2011}, and induce as a
consequence a flattening of the entropy profiles in cluster outskirts
and a bias in gas masses and hydrostatic masses
\citep[e.g.][]{Roncarelli_2006, Nagai_Lau_2011, Simionescu2011, 
Roncarelli_2013, Ettori_2013, Eckert_2013_1, Eckert_2013_2}.  
On the other hand, X-ray and SZ effect measurements are affected in 
different ways by gas density and pressure clumping. For instance, 
if the gas clumping is isobaric, the X-ray surface brightness will be 
augmented by a factor $n_e^{3/2}$, whereas the SZ effect will not be altered. 
On the contrary, pressure clumping could affect significantly the SZ 
scaling relations and the cosmological implications derived from the analysis 
of the SZ power spectrum \citep[e.g.][]{Battaglia_2014}.

The aim of this paper is to perform a detailed analysis of the ICM
thermal pressure distribution, the clumping factor and the SZ 
scaling relations of a sample of simulated galaxy clusters and groups 
that  has been shown to produce a realistic diversity between cool-core 
(CC) and non-cool-core (NCC) systems \citep[][]{Rasia_2015}.  
Different cluster properties are analysed as a function of redshift, ICM 
physics, cluster mass, cluster dynamical state and cluster cool-coreness.

This paper is organized as follows. In Section~\ref{sec:simulations}
we briefly introduce the main characteristics of our simulated
samples of galaxy clusters and groups. In Section~\ref{sec:pressure} we
analyze the radial pressure profiles of our sample of objects as a
function of mass, dynamical state, redshift and physics included in
our simulations.  Section \ref{sec:clumping} shows results on the ICM
clumping, both in density and pressure,
whereas Section \ref{sec:SZ} shows the corresponding SZ
scaling relations and their redshift evolution.  Finally, in Section
\ref{sec:summary} we summarize our main results.

\section{The simulations}
\label{sec:simulations}

\subsection{The simulation code}
\label{subsec:code}

We present the analysis of a set of hydrodynamic simulations of galaxy
clusters performed with an upgraded version of the TreePM--SPH code
{\footnotesize {\sc GADGET-3}} \citep[][]{springel05}.  The 
sample consists in zoomed-in simulations of 29 Lagrangian regions
extracted from a parent N-body cosmological simulation \citep[see][for
details on the initial conditions]{Bonafede2011}. The 29 regions
have been identified around 24 massive dark matter halos with
$M_{200}>8\times 10^{14}\, \msun$ and 5 less massive systems with
$M_{200}=[1-4]\times 10^{14}\, \msun$.
As described in \citet{Planelles_2013_b}, each of the 
low-resolution regions has been re-simulated, with improved resolution
and by including the baryonic component. 
The simulations consider a flat $\Lambda$CDM cosmology with
$\Omega_{\rm{m}} = 0.24$, $\Omega_{\rm{b}} = 0.04$,
$H_0=$~72~km~s$^{-1}$~Mpc$^{-1}$, $n_{\rm{s}}=0.96$, and $\sigma_8 =
0.8$.
For the DM particles the mass resolution is $m_{\rm{DM}} =
8.47\times10^8 \, \msun$, while the initial mass for the gas
particles is $m_{\rm{gas}} = 1.53\times10^8\, \msun$. 
Gravity in the high-resolution region
is calculated with a Plummer-equivalent softening length of $\epsilon =
3.75\, h^{-1}$ kpc for dark matter and gas   
and $\epsilon = 2\, h^{-1}$ kpc for black hole and stellar
particles. The softening is kept fixed in comoving coordinates for all
particles, except that, in the case of DM, it is given in physical units 
below $z=2$.

For the hydrodynamical description, we use an updated formulation of 
SPH as presented in \citet{Beck_2015}. This new SPH model has a number 
of features that improve its performance in contrast to standard SPH. These  
include higher-order interpolation kernels and derivative operators as 
well as advanced formulations for artificial viscosity and thermal diffusion. 
With these improvements, 
we have performed three different sets of  re-simulations characterized by 
different prescriptions for the physics of baryons:
\begin{itemize}
\item {\tt \nr}. Non-radiative hydrodynamical simulations based on the
  improved SPH formulation presented in \citet{Beck_2015}.

\item {\tt \w}. Hydrodynamical simulations accounting for the effects
  of radiative cooling, star formation, SN feedback and metal
  enrichment, as already explained in \citet{Planelles_2013_b}.  The
  chemical evolution model by \cite{tornatore_etal07} is employed to
  account for stellar evolution and metal enrichment. Metal-dependent
  radiative cooling rates and the effects of a uniform UV/X-ray
  background emission are included according to the models by
  \cite{wiersma_etal09} and \cite{haardt_madau01},
  respectively. Fifteen chemical species (H, He, C, Ca, O, N, Ne, Mg,
  S, Si, Fe, Na, Al, Ar, Ni) contribute to cooling. The star formation
  model is based on the original prescription by
  \cite{springel_hernquist03}, where galactic outflows originated by
  SN explosions are characterized by a wind velocity of $v_w=350\,
  km/s$.

\item {\tt \agn}. Hydrodynamical simulations accounting for identical
  physics as in the {\tt \w} case, plus AGN feedback.  
  The sub-grid model for super-massive black
  hole (SMBH) accretion and AGN feedback employed in our simulations
  has been presented in \citet{Steinborn_2015} and represents an
  improved version of the original model described in
  \cite{springel_etal2005}. Here, mechanical outflows and
  radiation are combined and included in the form of thermal
  feedback. The efficiency of radiative and mechanical feedback depend on both the
  (Eddington-limited) gas accretion rate and the SMBH mass, providing
  a smooth transition between radio and quasar mode. 
  We assume that a fraction $\epsilon_f=0.05$ of the radiative feedback couples to the 
  surrounding gas as thermal feedback.
  In addition, the accretion is computed separately for hot and cold gas. 
  In this work we neglect hot gas accretion ($\alpha_{\mathrm{hot}}=0$) and 
  consider only cold gas accretion, for which the Bondi accretion rate is boosted 
  by a factor $\alpha_{\mathrm{cold}}=100$ \citep[see][]{Gaspari_2015_b}.
  For further details on the model and its performance, we refer to the work by
  \citet{Steinborn_2015}.

  The set of {\tt AGN} simulations has been shown to generate, for the
  first time, a coexistence of CC and NCC simulated clusters with
  thermal and chemodynamical properties in good agreement with
  observations \citep[see][for further details]{Rasia_2015}.  
  This result is a consequence of the
  combined action of both the artificial conduction term included in
  the new hydro scheme and the new AGN feedback model, which improves
  the mixing capability of SPH and regulates the formation of stars
  in massive halos.

\end{itemize}

\begin{figure*}
 {\includegraphics[width=5cm]{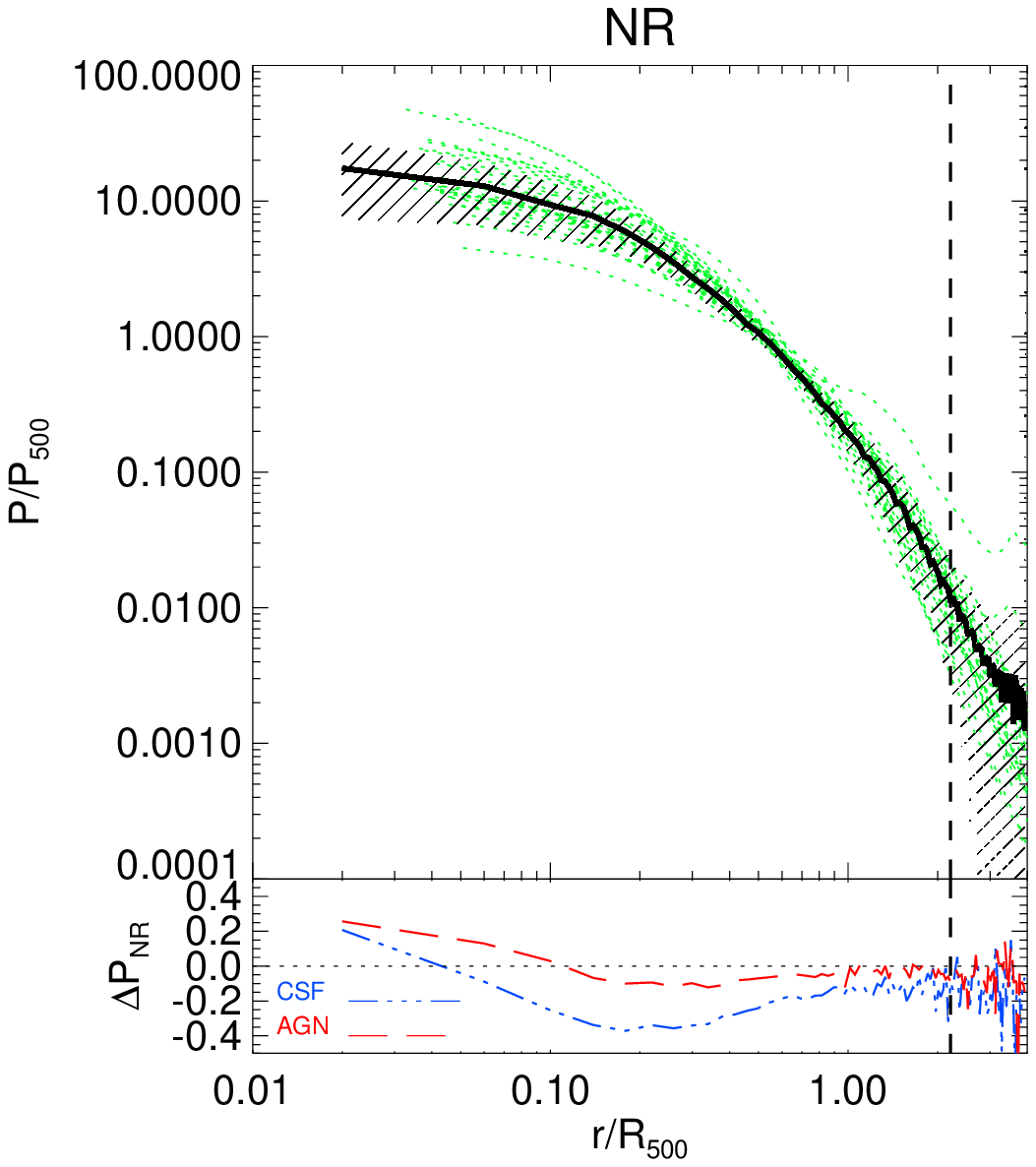}}
 \hspace{0.5cm}
 {\includegraphics[width=5cm]{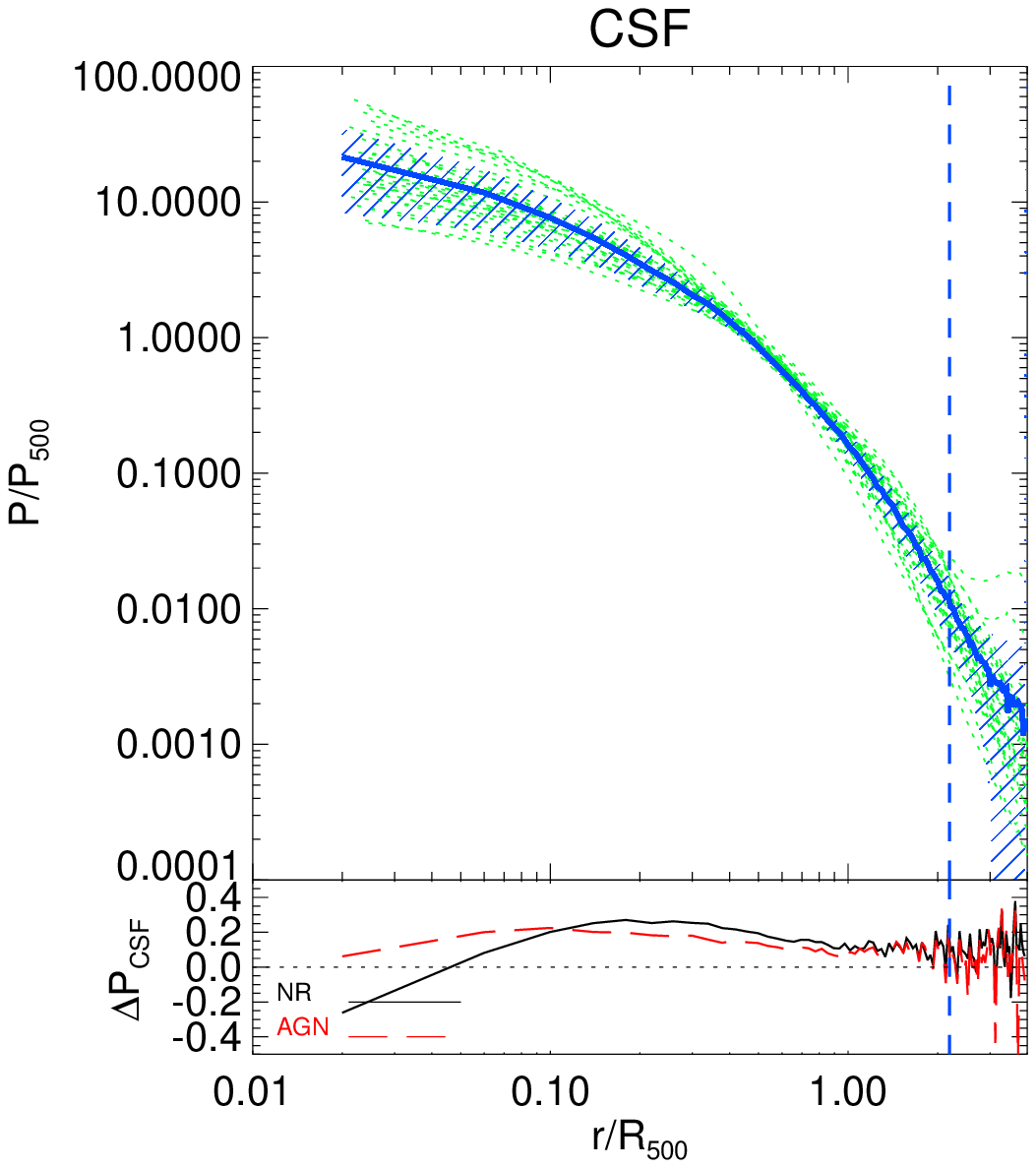}}
  \hspace{0.5cm}
 {\includegraphics[width=5cm]{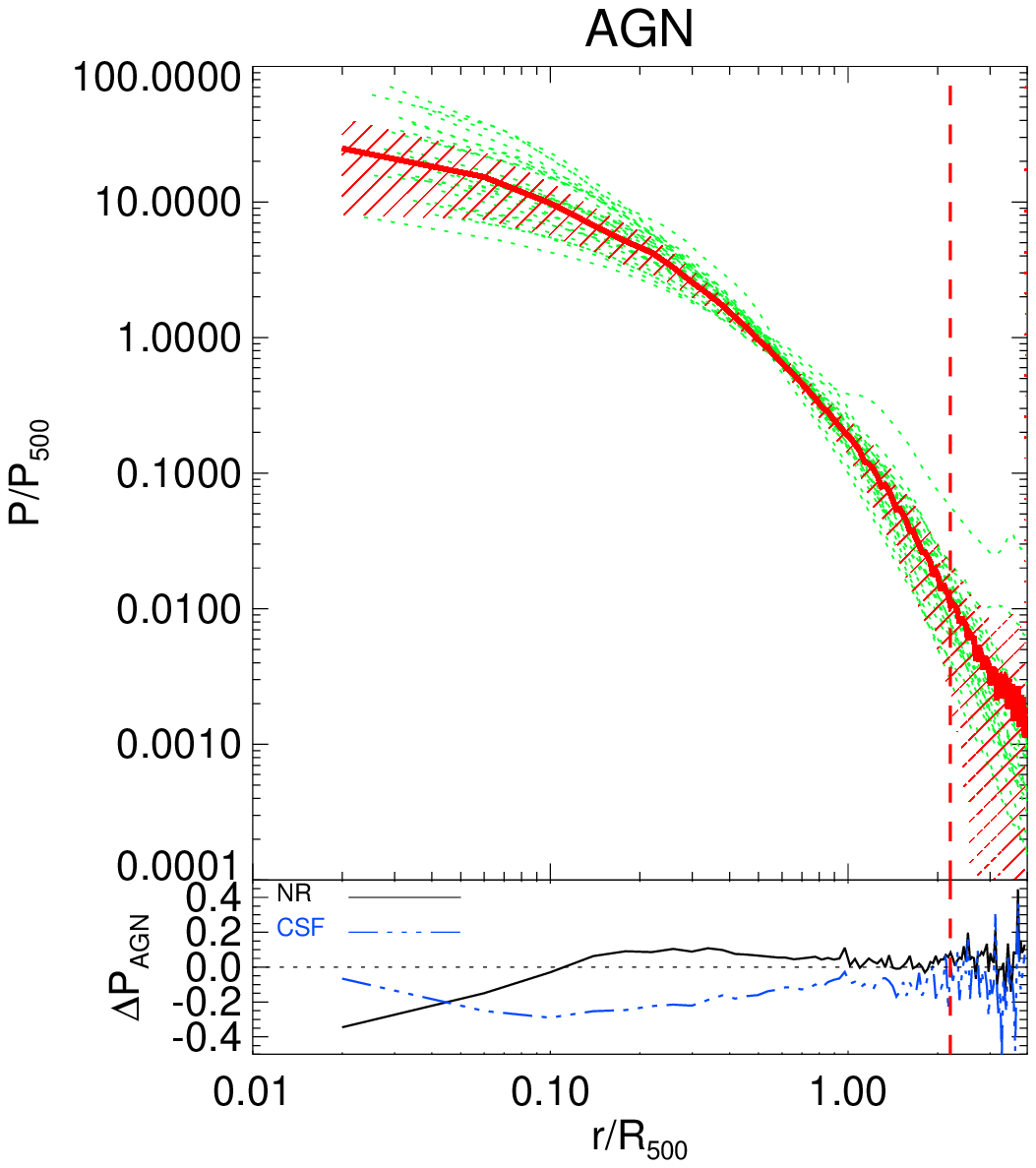}}
  \caption{Individual $P/P_{500}$ pressure profiles (green dotted
    lines) out to $4R_{500}$ at $z=0$ for our sample of 29
    central groups and clusters in the {\tt \nr}, {\tt \w} and {\tt
     \agn} simulations (panels from left to right, respectively).  In
    each panel, the continuous thick line represents the corresponding
    mean profile of each sample, whereas the streaky area stands for
    1-$\sigma$ scatter around the mean. The vertical line marks the mean
    value of $R_{vir}$, in units of $R_{500}$, within each sample.
    The relative difference between the mean profiles obtained for
     the reference model in each panel and the other two models is shown in the bottom 
    panels as $\Delta P_{REF}=(P_{mean}-P_{REF})/P_{mean}$.}
\label{fig:profiles_1}
\end{figure*}

\subsection{The sample of simulated clusters}
\label{subsec:sample}

We identify groups and clusters in the high-resolution regions
following a two-steps procedure.  First, we run a friends-of-friends
(FoF) algorithm with a linking length equal to 0.16 in units of the
mean separation of the high-resolution DM particles. This step
provides the centre of each halo, corresponding to the 
DM particle position within each FoF group with the minimum 
gravitational potential.  Then, a spherical overdensity algorithm is
applied to each halo and at each considered redshift in order to get
the radius $R_{\Delta}$ confining an average density equal to $\Delta$
times the corresponding critical cosmic density, $\rho_c(z)$.  In the
following, we will refer to an overdensity $\Delta=200, 500, 2500$, or to
the virial overdensity \citep{Bryan1998}.

Following this procedure, we identify, within each of our three sets
of re--simulations, a sample of $\sim 100$ clusters and
groups with $M_{500}>3\times 10^{13}h^{-1}M_\odot$ at $z=0$. In the
following, depending on the property to be analyzed, we will either
refer to the {\it ``reduced sample''} of 29 central halos (formed by 24 
clusters with masses $4\times 10^{14}h^{-1}M_\odot \leq M_{500} \leq
2\times 10^{15}h^{-1}M_\odot$ and 5 isolated groups with $6\times
10^{13}h^{-1}M_\odot \leq M_{500} \leq 3\times 10^{14}h^{-1}M_\odot$
at $z=0$) or to the {\it ``complete sample''} of $\sim 100$ systems 
(formed by all the halos found within each region in addition to the central
one). In general, we will use the reduced sample of groups and
clusters throughout the paper, whereas the whole sample will be
employed in the analysis of the scaling relations.
  
Based on different cluster properties, we have considered two
different classifications for the groups and clusters belonging to the
reduced sample in our {\tt \agn} simulations:
 \begin{itemize}
  
 \item CC-like and NCC-like halos. Following the criteria presented in
   \citet{Rasia_2015}, we classify our systems in the {\tt \agn}
   simulations according to their core thermodynamical properties. In
   particular, those systems with a central entropy $K_0< 60\, keV
   cm^{2}$ and a pseudo-entropy $\sigma<0.55$ are defined as CCs;
   those systems not satisfying these conditions are instead
   considered as NCC objects. With this criterion, 11 out of 29 halos are
   classified as CC clusters. As described in \citet{Rasia_2015}, the
   entropy and metallicity profiles of simulated CC and NCC
   objects are in good agreement with observations of these two
   cluster populations.
  
 \item Regular and disturbed halos. Groups and clusters are classified
   according to their global dynamical state. In order to do so, we
   combined two different methods generally used in simulations
   \citep[e.g., ][]{Neto_2007, Meneghetti_2014}: (a) the centre shift, given by the
   offset between the position of the cluster minimum potential and
   its centre of mass in units of its virial radius
   \citep[e.g.,][]{Crone_1996, Thomas_1998, Power_2012}, that is,
   $\delta r=||r_{min} - r_{cm}||/R_{vir}$; and (b) the mass fraction
   contributed by substructures, that is, $f_{sub}=M_{sub}/M_{tot}$,
   where $M_{sub}$ and $M_{tot}$ are, respectively, the total mass in
   substructures and the total mass of a given system as provided by
   SUBFIND \citep{Springel2001, Dolag2008}.  We consider a system to
   be regular when $\delta r < 0.07$ and $f_{sub} < 0.1$, whereas
   those systems with larger values for $\delta r $ and $f_{sub}$ are
   labeled as disturbed. Clusters for which both criteria are not
   concurrently satisfied are considered as intermediate
   cases. Following this classification we end up with 6 regular, 8
   disturbed and 15 intermediate systems \citep[see also][]{Biffi_2016}.
  
  \end{itemize}

\section{Pressure profiles}
\label{sec:pressure}

\subsection{Scaled pressure profiles}
\label{sec:pressure_1}

The radial pressure profile of each system is computed by assuming an
ideal gas equation of state.  Therefore, the volume-weighted estimate of the gas 
pressure is computed as
\begin{equation}
P\,=\,{\sum\limits_i^{} p_i dV_i\over \sum\limits_i^{} dV_i}\,,
\label{eq:pgen}
\end{equation}
where $p_i=(k_B/ \mu m_p)\rho_i T_i$, $dV_i$, $\rho_i$ and $T_i$ are
the contributions to pressure, volume, density and temperature,
respectively, of the $i$-th gas particle ($\mu$, $m_p$ and $k_B$ 
are the mean atomic weight, the proton mass and the Boltzmann
constant, respectively).
These individual pressure profiles have been computed using 100 
linearly equispaced bins within $2R_{vir}$, ensuring at least 100 gas 
particles within the innermost radial bin.

In order to compare with observational data, we use the dimensionless
pressure profiles, $p(x)=P(x)/P_{500}$, where $x=r/R_{500}$
and $P_{500}$, the `virial' characteristic pressure as provided by the HE 
condition \citep[e.g., see][]{Nagai_2007_2, Arnaud_2010}, can be written as
\begin{equation} 
  P_{500} = 1.45\times 10^{-11} \,{\rm erg \; cm^{-3}}\; \left(
    \frac{M_{500}}{10^{15}\,h^{-1}M_{\odot}}\right)^{2/3} E(z)^{8/3}
  \, , 
\label{eq:p500} 
\end{equation}
where $E^2(z)=\Omega_m(1+z)^3 + \Omega_{\Lambda}$.
Note that $P_{500}=n_{g,500}k_BT_{500}\propto (\mu_e/\mu) f_b M_{500}^{2/3}$.
The numerical coefficient in the r.h.s. of
Eq.~\ref{eq:p500} was derived by \citet{Nagai_2007_2} assuming
specific values for the baryon mass fraction $f_b$, the mean molecular
weight $\mu$ and the mean electronic molecular weight $\mu_e$.  In
order to make a proper comparison of the scaled pressure profiles from
different observational and/or simulated samples, 
in the following we will re-scale observed and simulated profiles to a common  value of $f_b$,
which introduces the main correction (although very small) in the computation of $P_{500}$.

Figure \ref{fig:profiles_1} shows the individual scaled pressure
profiles (green dotted lines) out to $4\times R_{500}$ at $z=0$ for
the reduced sample of systems in the {\tt \nr}, {\tt \w} and {\tt
\agn} simulations, respectively.  The mean scaled pressure profile
derived in each case is shown by a thick line together with
1-$\sigma$ scatter around it (streaky area).  Independently of the
physics included in our simulations, and despite the existence of 
some outliers, individual dimensionless pressure
profiles follow a nearly universal shape with a relatively small
scatter around the average, especially at intermediate radii,
$0.3\mincir r/R_{500}\mincir 1$, where we obtain a mean scatter of
$\sim 11$ per cent.  This result is in agreement with the
characteristic cluster-to-cluster dispersion at intermediate radii
obtained for the pressure profiles derived for both observed
\citep[$\sim 10-40$ per cent; e.g.][]{Arnaud_2010, Sun_2011} and
simulated clusters \citep[e.g.][]{Borgani_2004, Nagai_2007_2}.
Deviations from the self-similar behavior, mainly apparent in inner
($r\le 0.3R_{500}$) and outer clusters regions ($r\ge R_{500}$),
result from the combination of a number of factors, such as, the
effect of different feedback processes or the particular gas
distribution and dynamical structure of the considered systems.
Indeed, whereas the scatter in central regions ($r\mincir 0.2 /R_{500}$) 
is larger (by a factor of $\sim 1.5$) in the {\tt \agn} case,  
the fact that the three models show a similar 
scatter at larger radii indicates that the different formation histories of 
individual systems may also play a significant role. 
As we will discuss in Section \ref{sec:SZ}, given the relation between \yx\ and
\ysz\ and the obvious connection between cluster mass and pressure,
this low scatter explains why the $Y_X-M$ relation is usually observed
to be a good proxy for mass \citep[e.g.][]{Kravtsov_2006, Fabjan_2011,
Planelles_2013_b}.

In order to highlight the differences between the mean pressure profiles obtained 
for each model, we show in the bottom panels of  Fig.~\ref{fig:profiles_1} their relative 
difference computed  as $\Delta P_{REF}=(P_{mean}-P_{REF})/P_{mean}$, where 
$P_{REF}$ stands for the reference model in each panel (i.e., from left to right, {\tt \nr}, 
{\tt \w} and {\tt \agn}) and $P_{mean}$ stands for the other two models in comparison. 
From these panels it is clear that, for $r\mincir 0.1R_{500}$,  the {\tt \agn} model produces the 
higher central pressure (by $\sim 20-30$ per cent) while it is more similar 
to the {\tt \nr} run than to the  {\tt \w} at larger radii.

We have checked that,  the mean and the median radial
profiles of our sample of clusters are very similar to each other.
Therefore, in order to compare with observational data and unless
otherwise stated, in the following we will use the mean simulated
profiles shown in Fig.~\ref{fig:profiles_1}.

\begin{figure*}
\begin{center}
{\includegraphics[width=8cm]{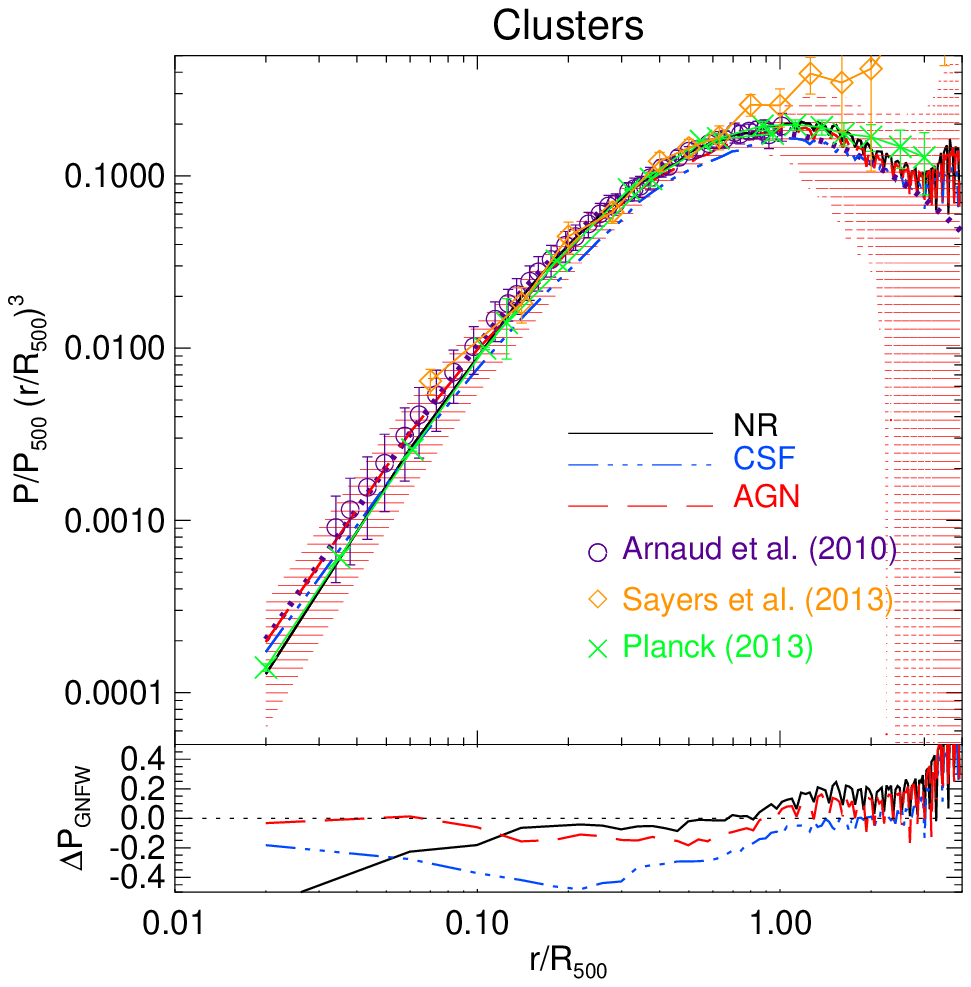}}
{\includegraphics[width=8cm]{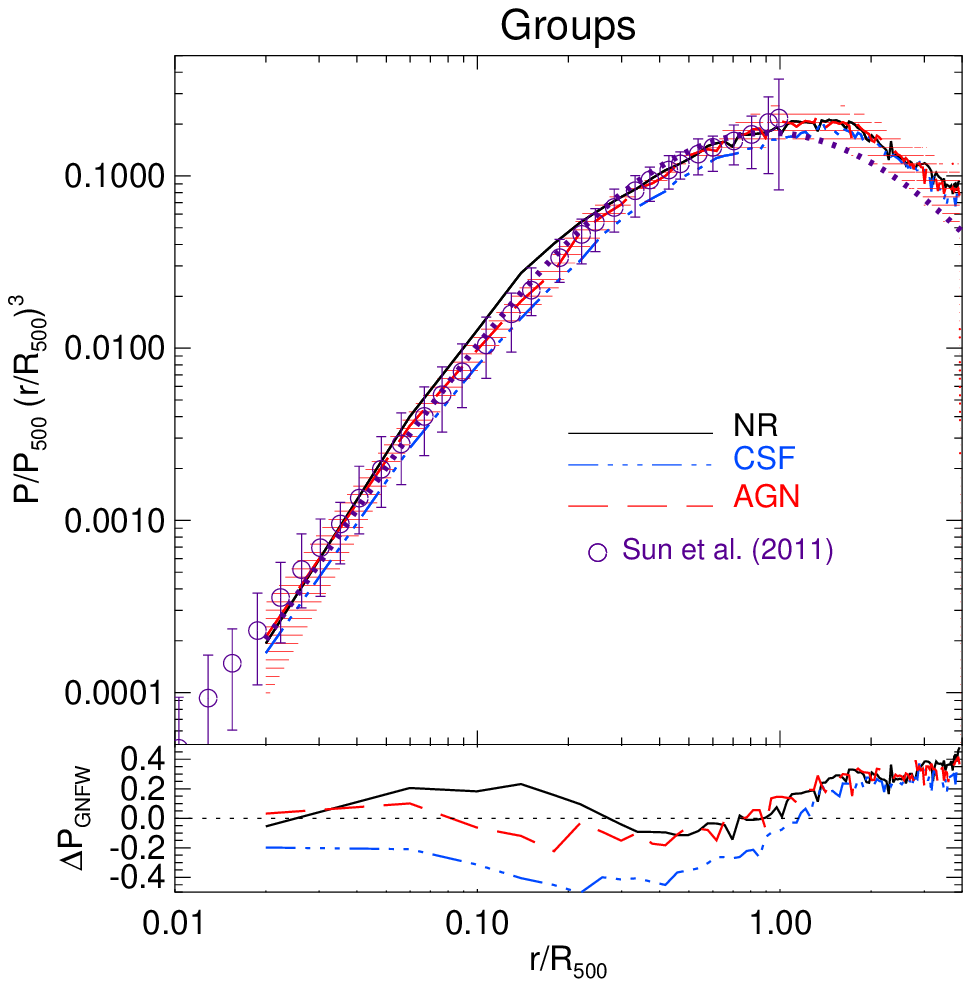}}
\end{center}
\caption{Mean $P/P_{500}$ pressure profiles, weighted by
  $(r/R_{500})^3$, out to $4\times R_{500}$ at $z=0$ for our sample of
  24 central clusters (left panel) and 5 central groups (right panel).
  Black, blue and red lines stand for the mean profiles in the {\tt
  \nr}, {\tt \w} and {\tt \agn} simulations, respectively.   
  For the sake of clarity, the streaky red area shows 1-$\sigma$ deviation around the
  mean profile in the {\tt \agn} run.
  Observed pressure profiles from \citet{Arnaud_2010}, \citet{Sun_2011},
  \citet{Planck_2013} and \citet{Sayers_2013} are used for comparison
  (different symbols with error bars). The purple dotted line shown in 
  both panels corresponds to the best-fit profile to a GNFW model
  as obtained by \citet{Arnaud_2010} (see Section \ref{sec:fit} for further details). 
  The relative difference between the simulated profiles and the best-fit profile  
   of \citet{Arnaud_2010} is shown in the bottom panels.}
\label{fig:different_masses}
\end{figure*} 

\subsection{Comparison with observational samples}

In this Section we will compare our simulated data with 5 different 
observational samples briefly described as follows.

\citet[][]{Arnaud_2010} presented the analysis of
the pressure profiles of the \rexcess\ sample \citep{Bohringer_2007},
a selection of 33 nearby ($z<0.2$) clusters within a mass range of
$10^{14}M_\odot<M_{500}<10^{15}M_\odot$ observed with \xmm.  They
derived a universal pressure profile out to cluster edges by combining
observations within the radial range $[0.03-1]R_{500}$ with
results from numerical simulations within $[1-4]R_{500}$.  According
to the criteria defined in \citet{Pratt_2009}, they divided this
sample of clusters in cool core (those with peaked density profiles in
central regions) and morphologically disturbed (those with a large
centre shift parameter) systems. The mass $M_{500}$ of each cluster was
estimated using the $M_{500}-Y_X$ scaling relation as obtained by
\citet{Arnaud2007}.

\citet[][]{Planck_2013} analysed the SZ pressure profiles of a sample
of 62 local ($z<0.5$) massive clusters with
$2\times10^{14}M_\odot<M_{500}<2\times10^{15}M_\odot$, identified from
the Planck all--sky survey.  These SZ-selected systems,
taken from the \planck\ early SZ sample \citep{Planck2011_c}, have been
also followed-up with \xmm\ observations.  The radial SZ effect signal of
this sample of clusters was detected out to $3R_{500}$. In order to
provide a proper radial analysis, they combined this data with \xmm\
observed profiles within $[0.1-1]R_{500}$. Masses and radii of these
clusters were derived by means of the $Y_X-M_{500}$ relation by
\citet{Arnaud_2010}.  Following the classification criterion described
in \citet{Planck_2011b}, they divided the whole sample of clusters in
22 CC and 40 NCC systems.

\citet[][]{Sayers_2013} analyzed the pressure profiles of a sample of
45 galaxy clusters, with a median mass of
$M_{500}=9\times10^{14}M_\odot$, within a redshift range of $0.15
\mincir z \mincir 0.89$. Total masses were derived using the gas mass as a proxy 
\citep[see][for details]{Mantz_2010}. These clusters, imaged by means of SZ
observations with the Bolocam at the Caltech Submillimeter
Observatory, span a radial range within $[0.07-3.5] R_{500}$.  The
whole sample of clusters is formed by  17 CC
clusters (defined according to their X-ray luminosity
ratio), 16 disturbed objects (defined according to their X-ray centroid shift), 
and 2 systems that are disturbed and have a CC.

\citet[][]{McDonald_2014} presented the X-ray analysis of a sample of
80 galaxy clusters from the 2500 $deg^2$ SPT survey.  
The masses of the systems, with $M_{500}\magcir3\times10^{14}E(z)^{-1}M_{\odot}$,
were derived using the $Y_X-M$ relation by \citet{Vikhlinin_2009}. 
Assuming a self-similar temperature profile, they made an X-ray fit to their
clusters, which were divided in different subsamples according to
their redshift and central densities. In this way, they constrained
the shape and evolution of the temperature, entropy and pressure
radial profiles within $1.5R_{500}$ and out to $z=1.2$.
 
As for the observation of smaller systems, \citet{Sun2009, Sun_2011} analyzed
\chandra\ archival data for 43 nearby (0.012 $<z<$ 0.12) galaxy groups
with $M_{500} = 10^{13} - 10^{14} h^{-1}$ M$_{\odot}$.  They obtained
gas properties out to at least $R_{2500}$ for the whole sample
and out to $R_{500}$ for 11 groups. For an additional subsample of 12
groups, they extrapolated to derive properties at $R_{500}$. 
\vspace{0.3cm} 
 
It is important to note the diversity of these sets of observations
both in their nature (i.e., X-rays combined with simulations, X-rays
or SZ observations) and in the properties of the considered
cluster/group samples.  In addition, there is also a difference in the
criteria used to classify the observed clusters in CC and NCC systems.
Given the diversity of these observational samples, the comparison
with the pressure profiles of our simulated systems will help us to
provide a more robust analysis of the existing dependencies on cluster
mass, redshift and dynamical state.

We note that, when cluster masses are derived assuming HE, 
as it is done in most X-ray analyses, they are 
underestimated by a factor $\sim10-20$ per cent \citep[e.g.][]{Arnaud_2010}.
In order to compare simulated and observed pressure profiles,  given the dependencies 
of $P_{500}$ and $R_{500}$ on mass, the hydrostatic mass
bias would imply a reduction in these quantities, translating, therefore, the scaled profiles. 
Since the change induced by this bias is relatively small  \citep[e.g., a $15$ per cent  mass bias implies a $\sim 10$ 
per cent change in $P/P_{500}$ and $\sim 5$ per cent in $R_{500}$;][]{Planck_2013}, given that its magnitude 
is still uncertain, we decide not to account for this correction and refer the reader to more detailed analyses
that specifically address the issue of HE \citep[e.g][and references therein]{Biffi_2016}.

In addition, the definition of $P_{500}$ given by Eq.~\ref{eq:p500} reflects the
variation of the pressure within $R_{500}$ with mass and redshift as
predicted by the standard self--similar model. Unlike  \citet{Arnaud_2010} or \citet{Planck_2013}, 
but in a similar way to \citet{Sayers_2013}, we do not include any additional correction to the mass 
dependence of $P_{500}$ \citep[see equations 7-8 of][]{Arnaud_2010}. We have checked that, for the 
sample of AGN systems considered here, the $M_{500}-Y_{X, 500}$ scaling relation is self-similar out to 
redshift 1 and, therefore, there is no need to introduce any additional non-self-similar mass dependence  
\citep[as shown in][a similar result is also obtained for our complete set of systems in our three models]{Truong_2016}.

\begin{figure*}
{\includegraphics[width=8cm]{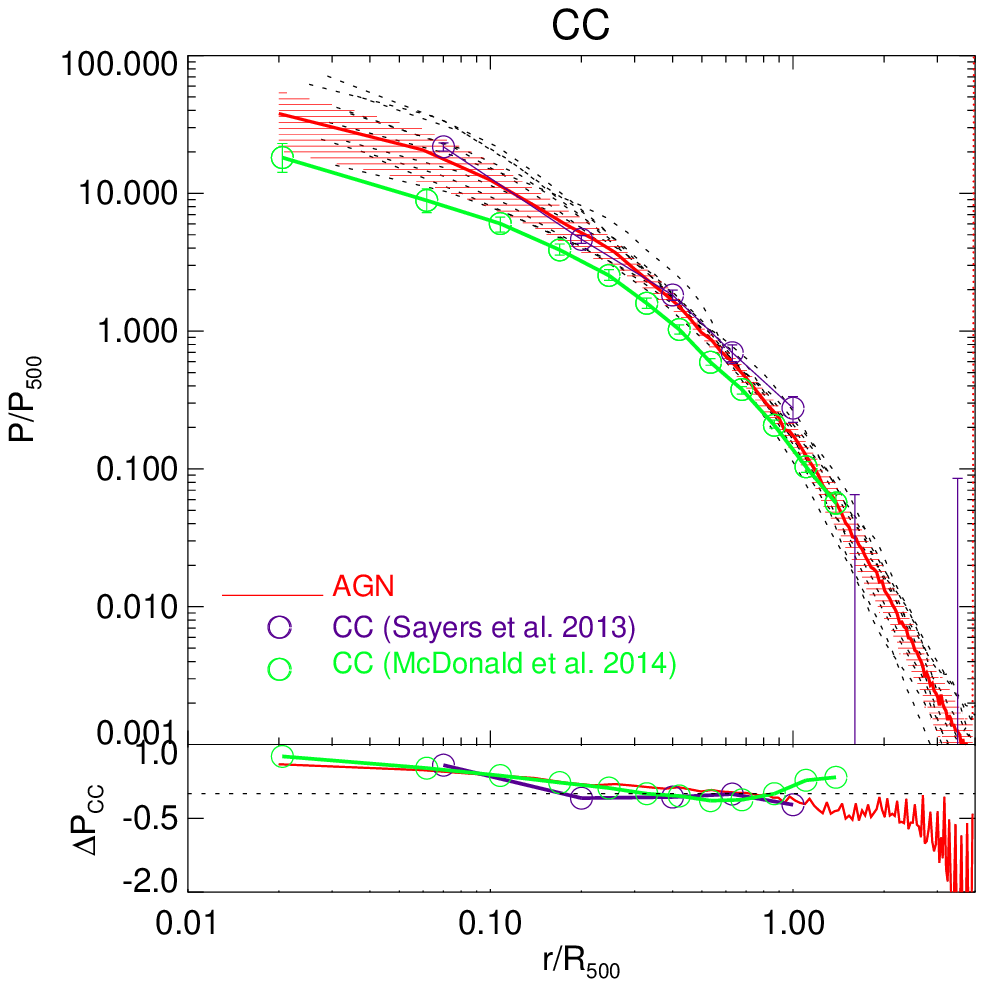}}
{\includegraphics[width=8cm]{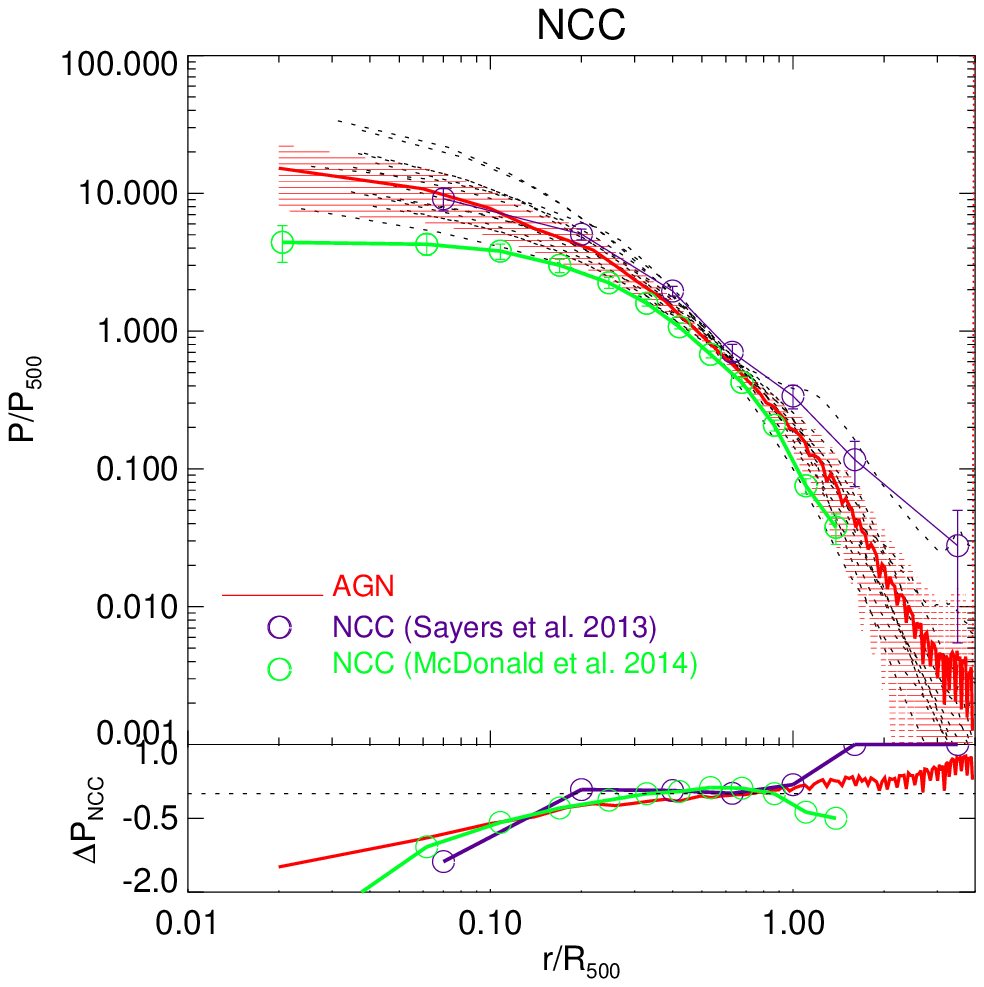}}
\caption{Mean radial profiles of scaled pressure, $P/P_{500}$, out to
  $4 R_{500}$ at $z=0$ for our CC (left panel) and NCC (right panel)
  clusters within our {\tt \agn} simulations.  Individual profiles are
  shown with thin dotted black lines. The mean profile of each sample
  is represented instead by a continuous red line surrounded by a
  shaded area representing 1-$\sigma$ scatter around the mean
  profile.
  The average pressure profiles of observed CC and NCC clusters as derived 
  by \citet{Sayers_2013} (purple open circles with error bars) and 
  \citet{McDonald_2014} (green open circles with error bars) are used for comparison.
  In the bottom panels we show the relative difference between the mean profiles of CC and NCC clusters
  as obtained in all the samples. In particular,  $\Delta P_{CC}=[(P/P_{500})_{CC}-(P/P_{500})_{NCC}]/(P/P_{500})_{CC}$ 
  ($\Delta P_{NCC}$ is computed in a similar way). The horizontal dotted line marks a zero relative difference.
  }
\label{fig:profiles_cc_ncc}
\end{figure*}

\subsubsection{Dependence on cluster mass and physics included}
\label{sec:massdepen}

The left and right panels of Fig.~\ref{fig:different_masses} show,
respectively, the mean pressure profiles $P/P_{500}$, scaled by
$(r/R_{500})^3$, for the sample of 24 central clusters and 5 central
groups within each of our simulations.  We compare these mean profiles
with different observational samples of groups and clusters.  
For completeness, we also compare with the
universal pressure profile obtained by \citet{Arnaud_2010}.
The relative difference between the mean simulated profiles and the
best-fit to a GNFW model by \citet{Arnaud_2010}
is shown in the bottom panels as $\Delta P_{GNFW}=(P_{mean}-P_{fit})/P_{mean}$.

We focus first on the analysis of the pressure profiles  of massive 
clusters and their dependencies on the different
physical models included in our simulations (left panel of
Fig.~\ref{fig:different_masses}). In general, simulated and observed
profiles show a good agreement (within 1-$\sigma$) and a considerably small scatter at
intermediate cluster radii ($0.2\mincir r/R_{500}\mincir 1$),
indicating that they are nearly self-similar in this radial range.
Within inner regions, the {\tt \agn} simulations tend to produce a
slightly higher central pressure (by a factor $\sim1.2-1.5$) 
than the other two models, whereas
the {\tt \nr} runs produce the lowest values. Nevertheless, all our
simulated sets tend to produce results that are close to each other
and in remarkably good agreement with observations, especially to
those from the REXCESS sample \citep{Arnaud_2010}. Despite the
consistency with the data, the dispersion of the profiles in these
internal regions is relatively large, presumably due to both the
different formation histories and the different effect that radiative
cooling and AGN feedback play on the ICM thermal properties.  Indeed,
thanks to the improved SPH scheme employed in our
simulations, the {\tt \nr} model produces a lower central pressure in
clusters and a smoother thermal distribution than the standard SPH.
Quite interestingly, the combination of the new hydro scheme and the
AGN feedback model, allows the {\tt \agn} simulations to efficiently
compensate overcooling and to keep pressurized a large amount of
low-entropy gas that now remains in the hot phase, providing therefore
a higher pressure in the center.
In outer cluster regions ($r \magcir R_{500}$), however, our three
simulation sets produce quite similar (within a few per cent) 
mean profiles  and the scatter between them is slightly reduced.  
The dispersion around the mean
profiles is however larger as a result of the additional matter and
dynamical activity characterising these external cluster regions.  In
general, even when the agreement with the observed profiles is quite
encouraging, simulated profiles are in better agreement with the
observed SZ mean profile from \citet{Planck_2013}, showing as well a
larger cluster-to-cluster scatter than at intermediate regions. In
these outer regions, in agreement with previous studies
\citep[e.g.][]{Battaglia_2010}, our mean profiles are systematically
higher (by $\sim20$ per cent) than the universal profile presented in
\citet{Arnaud_2010}. Nevertheless, we would like to point out that 
we perfectly reproduce the outer part of the profile given by 
\citet{Arnaud_2010} when we restrict our sample  to the $\sim 20$ 
most massive clusters with $M_{500}\geq 6\times 10^{14}\, \msun$.

To further investigate the impact of the different physical models on
systems of different masses, we focus now on the analysis of  groups
(right panel of Fig.~\ref{fig:different_masses}). 
The combined action of different feedback
processes on systems of different masses is expected to affect in a
different way the distribution of pressure within the considered
systems, both in the core and in the outskirts. In our case, however,
independently of the physics included, groups and clusters show very
similar mean pressure profiles, suggesting that they are nearly
self-similar in all models.  This result seems to be in contradiction
with the effect seen in recent simulations
\citep[e.g.][]{Battaglia_2010, McCarthy_2014}, where the pressure
profiles of groups within radiative simulations are lower (higher) in
central (outer) regions than in more massive clusters, 
being in line with the expectation that feedback is more efficient in the
central regions of low-mass systems. Again, this apparent
contradiction disappears by selecting a sample of lower-mass groups.
Indeed, if we use our whole sample of systems (not only the 29 central
objects), we can set a lower mass threshold for the sample of groups,
that is, we can consider all the systems with $3.0\times 10^{13} \msun
\le M_{500} < 3.0\times 10^{14} \msun $. For this larger sample of
smaller groups, formed by more than 70 systems with a mean mass 
$\langle M_{500}\rangle\sim8.2\times10^{13} \msun$ (a lower value than
the mean mass of our sample of 5 isolated groups, that is, 
$\langle M_{500}\rangle\sim1.9\times10^{14} \msun$), both of our radiative
simulations produce, at $r\mincir R_{500}$, a significant decrease of
the pressure with respect to the non-radiative case (as an example, 
at $r=0.1R_{500}$, the {\tt \w} and {\tt \agn} models show a pressure lower by a factor of 
$\sim$ 2 and 2.7, respectively, than the {\tt \nr} run).  This reduction
of the central pressure induced by AGN feedback at the scale of groups
is in broad agreement with previous simulated results considering a 
similar range of masses \citep[e.g.][]{McCarthy_2014}.  Contrarily,
this effect is not so pronounced for a larger sample of high--mass
systems (a sample of more than 30 clusters with $M_{500} \ge 3.0\times
10^{14} E(z)^{-1} M_\odot$ and 
$\langle M_{500}\rangle\sim7.7\times10^{14} \msun$) for which the 
central pressure seems to be
quite insensitive to the inclusion of AGN feedback since both {\tt \w}
and {\tt \agn} simulations produce very similar profiles.  This
result, in agreement with recent findings from simulations
\citep[e.g.,][]{Planelles_2013_b, McCarthy_2014, Pike_2014}, is also
in line with the expectation that the total thermal content of the ICM
is only weakly affected by feedback sources in massive systems while
being more sensitive to feedback in less massive objects \citep[e.g.][]{Gaspari_2014}.  
In any case, it is important to point out that the inclusion of AGN feedback
in our simulations produces pressure profiles in galaxy groups that
are in good agreement  (within $\sim 15-20\%$) with the observations reported by
\citet{Sun_2011} and with the universal fit of \citet{Arnaud_2010}.

As for the outskirts of groups and clusters the mean pattern is
also quite similar. However, in the case of clusters, the scatter
increases progressively with radius being much larger than in groups.  
This increase in outer regions
is partially associated to the fact that, whereas our sample of clusters is formed
by 24 non-isolated massive systems, our sample of groups is composed by
5 isolated objects. Indeed, in the case of groups, the dispersion around the mean profile is quite
sensitive to the group selection, not only in terms of mass but also
depending on the environment where groups reside: when we
consider our larger sample of non-isolated groups, the dispersion also
increases significantly in the outermost regions.
In addition, there should be also a contribution from the infall of pressurized clumps or by gas in
filaments or stripped from merging halos.  As we will discuss below, this
is consistent with the higher degree of clumpiness observed in the
peripheral regions of massive clusters.  
Moreover, as suggested by recent studies \citep{Avestruz_2016}, non-thermal pressure, associated to gas motions 
in outer cluster regions, plays a major role in the ICM thermal evolution and, therefore, 
it can also affect the distribution of thermal pressure.

\subsubsection{Dependence on cool-coreness}
\label{sec:dynam}

\begin{figure*}
\begin{center}
\scalebox{1.4}{\includegraphics[width=12cm]{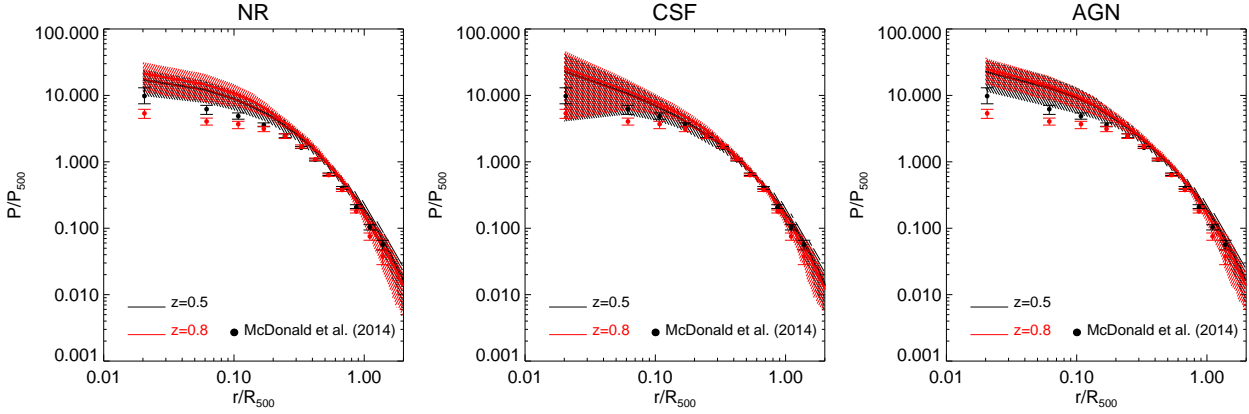}}
\end{center}
\caption{Mean pressure profiles for our sample of 29 main groups and
  clusters at $z=0.5$ (black line) and at $z=0.8$ (red line) within
  the {\tt \nr} (left panel), {\tt \w} (middle panel) and {\tt \agn}
  (right panel) simulations.  Colored regions in black and red
  represent, respectively, 1-$\sigma$ scatter around the mean profiles
  at $z=0.5$ and at $z=0.8$. Filled circles with error bars refer to
  the observational sample by \citet{McDonald_2014} at a mean redshift 
  of $0.46$ (black) and $0.82$ (red).}
\label{fig:against_obs_z}
\end{figure*}

Figure \ref{fig:profiles_cc_ncc} shows the mean pressure profiles
$P/P_{500}$ obtained for the subsamples of CC (left panel) and NCC
(right panel) clusters at $z=0$ within our {\tt \agn} simulations.
In order to compare with observations, the de-projected pressure
profiles of the disturbed and cool-core clusters of the BoXSZ sample,
as derived by \citet{Sayers_2013}, together with those derived by
\citet{McDonald_2014} for a sample of 80-SPT selected clusters are
also shown.  We note that these observed profiles are not exactly
at $z=0$ but at a median redshift of $0.42$ in the case of
\citet{Sayers_2013} and at $0.3< z< 0.6$ in the case of
\citet{McDonald_2014}\footnote{In order to correct a small calibration issue, the values of $r/R_{500}$ of the pressure profiles of this sample, shown in Figs.~\ref{fig:profiles_cc_ncc} and \ref{fig:against_obs_z}, have been multiplied by a factor 1.027 (M. McDonald, private communication).}. However, as we will discuss in 
Section \ref{sec:zevol}, since cluster pressure profiles do not show a 
noticeable z-evolution, we can perform this comparison.
Moreover, we remind that, given the different criteria used to classify clusters in CC/NCC, this comparison has to be intended as a qualitative approach to emphasize the similarity between the trend of the two classes.

As shown in the bottom panels of Fig.~\ref{fig:profiles_cc_ncc},
the mean profiles we obtain for the CC/NCC cluster
populations are in good agreement with each other at intermediate 
($0.2R_{500}<r<R_{500}$) radii. This is consistent with the work by \citet{Rasia_2015} as well as with a
number of observational results \citep[e.g.][]{Arnaud_2010, Planck_2013}.
However, as expected, these profiles clearly
diverge from each other in inner cluster regions ($r/R_{500}\mincir 0.2$), 
where observations generally report slightly higher
values of pressure and steeper profiles in CC than in NCC clusters.
The mean values of the core pressure and the shape of the profiles
obtained for our samples of CC and NCC systems are in good agreement
with the observational trend (at the innermost radius, our CC clusters show 
a mean pressure $\sim2.5$ times higher than our NCC clusters).  
In general, our results for these two
cluster populations show a good consistency (within 1-$\sigma$) 
with the observations by \citet{Sayers_2013}, especially in inner cluster regions. 
However, especially for NCC systems, our mean pressure profile in cluster
outskirts ($r>R_{500}$) is lower (e.g., at $r=3R_{500}$ by a 
factor of $\sim7$) than the one reported by \citet{Sayers_2013}, being in better 
agreement with the observational
determinations by \citet{Arnaud_2010} or \citet{Planck_2013} (see
\citealt{Sayers_2013} or \citealt{McDonald_2014}, for a more detailed
discussion).

Figure \ref{fig:profiles_cc_ncc} also shows that the mean profile
obtained for our sample of NCC objects has a higher dispersion,
especially in the outer regions, than for the sample of CC clusters. The
same happens between the different observational samples used for
comparison. As we will discuss in Section \ref{sec:clumping}, 
the larger dispersion in cluster outskirts is mostly produced by a clumpier gas 
distribution in outer regions of unrelaxed massive galaxy clusters.  In addition, as we will also see
in Section~\ref{sec:SZ}, small changes in the pressure profiles
depending on either the dynamical state or the cool-coreness of the
considered systems have a different impact on the corresponding \ysz$-M$
relation (see Table \ref{t:fit_YM}).
 
\subsection{Evolution with redshift}
\label{sec:zevol}

Thanks to its redshift independence, the SZ effect provides the ideal means
to trace the evolution of the ICM properties, especially in the outer
cluster regions. The young dynamical age of these regions make them
quite interesting to trace the recent history of cluster
assembly. However, they are hardly accessible to X--ray observations
of distant clusters because of surface brightness dimming.

Figure \ref{fig:against_obs_z} shows the mean pressure profiles
obtained for our sample of 29 central systems at $z=0.5$ and $z=0.8$
within each of our three sets of simulations. To compare with
observational data, we use the high-redshift ($\langle z\rangle=0.82$) and 
low-redshift ($\langle z\rangle=0.46$) pressure profiles  from the SPT sample 
presented in \citet{McDonald_2014}.  
Despite the differences in mass of the samples in this comparison,
independently of the considered redshift and physical model, we obtain
a relatively good agreement with the observational data throughout the
radial range.
The major differences arise at $r\leq 0.3 R_{500}$, where the {\tt \w} model 
shows the best agreement (within 1-$\sigma$ at both redshifts) with the data. 
In particular, simulations predict gas pressure in excess of the observed one in core
regions, $r\leq 0.1R_{500}$, with the {\tt \nr} and {\tt \agn} models 
showing the largest deviations (by $\sim$3- and $\sim$2.5-$\sigma$, respectively)
from the data at $z=0.8$.
Simulations are more consistent with the
observed values at intermediate radii, $0.1\leq r/R_{500} \leq 0.3$, 
where, again, the {\tt \nr} and {\tt \agn} runs show the largest deviation 
(by $\sim$2.5- and $\sim$2-$\sigma$, respectively) from the high-redshift data.
We note that part of the discrepancy in the normalization is due to the 
fact of comparing with the data by \citet{McDonald_2014}, since it shows  
lower pressure values than the rest of observational samples we have used 
for comparison (see Figs.~\ref{fig:different_masses}-\ref{fig:profiles_cc_ncc}). 
However, if we focus on the z-evolution, simulated and observed profiles show 
a similar trend.
Interestingly, if we compare with the {\tt \nr} model, including AGN physics
slightly increases the central pressure at both redshifts. This is due to the 
effect that AGN feedback and thermal conduction have in pressurizing the gas in 
cluster cores.
The agreement with observations improves, for all the three models, in outer regions, i.e. at
$r\ge 0.3R_{500}$,  being particularly encouraging at $z=0.5$, whereas at $z=0.8$ there are more 
apparent deviations.  In any case, in these external regions, observed and simulated profiles are
consistent within 1-$\sigma$ with each other.

As pointed it out by \citet{McDonald_2014}, from the analysis of the pressure profiles 
at $z=0.8$, we see that both observed and simulated profiles show a small but
measurable steepening in outer regions. This steepening can be due to
the existence of dense gas clumps which would tend to increase both
the pressure measurements and the scatter around the mean universal
profile. Moreover, the lower pressure that characterizes  the
outskirts at higher redshifts may indicate a delayed effect of heating
sources.

Overall, simulated and observed scaled pressure profiles show little
(if any) redshift evolution.  This lack of evolution suggests that any
non-gravitational processes should have played an action on ICM
pressure at even earlier epochs.  As shown in \citet{McDonald_2014},
our results on the evolution of pressure profiles are also broadly
consistent with previous simulations \citep[e.g.][]{Battaglia_2012a}
as well as with additional observations of high-redshift clusters
\cite[e.g.][]{Adam_2014}.

\subsection{Universality of pressure profiles}
\label{sec:fit}

\begin{table*}
  \caption{Values of the best-fit parameters describing the GNFW pressure profile given by Eq.~\ref{eq:pgnfw} for
    different sets of fixed parameters (indicated with $\dagger$). Results are shown
    for our sample of 24 massive clusters in the {\tt \nr}, {\tt \w} and {\tt \agn} simulations. 
    Results obtained for the subsamples of CC and NCC systems within the {\tt \agn} model are also shown.
    Fitted parameters in each case are shown with errors representing
    1-$\sigma$ uncertainty.}
\label{t:fit_gnfw}
\hspace*{0.3cm}{\footnotesize }
\begin{tabular}{@{}lcccccccc}
\hline

Simulation & Sample & z & $P_0$ & $c_{500}$ & $\gamma$  & $\alpha$ & $\beta$ \\ 
\hline
{\tt \nr}  &  Reduced  & $0.0$  &     $6.85\pm1.64$    &   $1.09\pm0.50$   &   $0.31^\dagger$    &    $1.07\pm0.28$     &   $5.46\pm1.15$  \\
     \\
{\tt \w}    & Reduced   & $0.0$  &   $6.35\pm1.89$    &   $0.63\pm0.52$   &   $0.31^\dagger$    &    $0.86\pm0.24$     &   $6.73\pm1.54$   \\
     \\
{\tt \agn}     & Reduced      & $0.0$  &     $8.25\pm2.90$    &   $0.54\pm0.54$   &   $0.31^\dagger$    &    $0.81\pm0.22$        &   $7.32\pm1.91$   \\
                  & Reduced-CC   & $0.0$  &   $11.96\pm3.59$    &   $0.64\pm0.44$   &   $0.31^\dagger$    &    $0.81\pm0.16$   &   $7.06\pm1.72$     \\
                  & Reduced-NCC   & $0.0$  &  $5.71\pm1.45$    &   $0.71\pm0.55$   &   $0.31^\dagger $    &    $0.93\pm0.22$   &   $6.34\pm2.14$   \\                  
     \\
     &   Reduced      & $0.0$  &             $3.99\pm5.84$    &   $1.19\pm0.14$   &   $0.52\pm0.22$    &    $1.17\pm0.15$    &   $5.0^\dagger$  \\
     &                       & $0.3$  &             $2.72\pm6.16$    &   $0.98\pm0.12$   &   $0.63\pm0.21$    &    $1.12\pm0.13$ &   $5.0^\dagger$    \\
     &                       & $0.5$  &             $3.46\pm6.06$    &   $1.07\pm0.12$   &   $0.54\pm0.21$    &    $1.12\pm0.15$    &   $5.0^\dagger$ \\
     &                       & $0.8$  &             $15.34\pm6.16$    &   $1.39\pm0.12$   &   $0.18\pm0.17$    &   $0.92\pm0.12$  &   $5.0^\dagger$   \\
     &                       & $1.0$  &             $24.19\pm5.28$    &   $1.24\pm0.08$   &   $0.14\pm0.12$    &   $0.77\pm0.04$   &   $5.0^\dagger$  \\
\hline
\end{tabular}

\label{tab:fittings}
\end{table*}

\begin{figure*}
\scalebox{0.8}{\includegraphics[width=21cm]{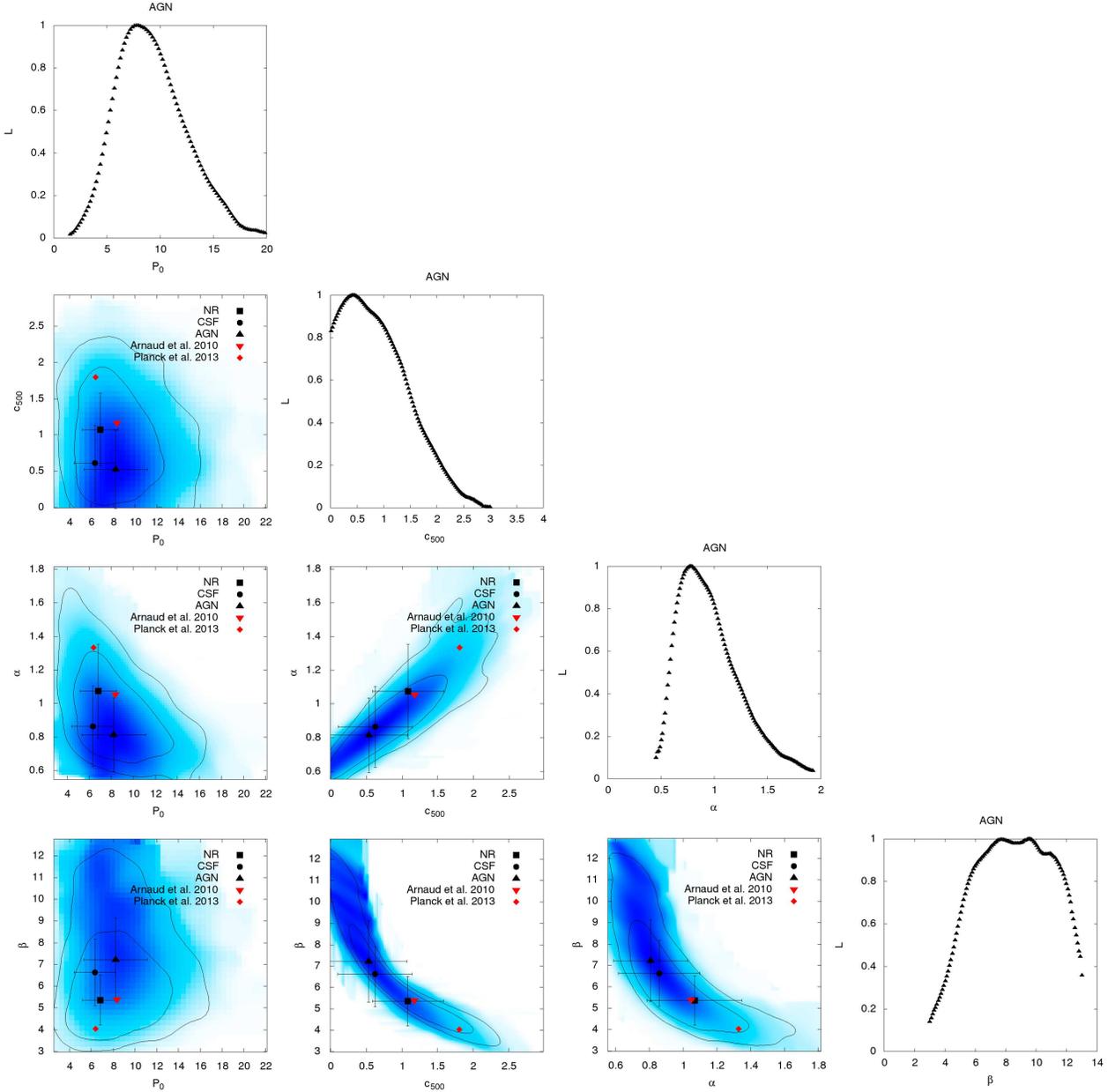}}
\caption{Likelihood confidence regions for the parameters of the GNFW model with $\gamma=0.31$ obtained for the sample of 
clusters within our {\tt \agn} simulations.
Solid lines on the 2D distributions represent the $68$ and $95$ per cent confidence levels.
Filled symbols in black (squares, circles and triangles, respectively) mark the best-fit values obtained for the 
{\tt \nr}, {\tt \w} and {\tt \agn} runs, whereas 
red symbols are for the best-fit values obtained by \citet{Arnaud_2010} and \citet{Planck_2013}.}
\label{fig:contours_cosmomc}
\end{figure*} 

\begin{figure}
\hspace*{0.8cm}
{\includegraphics[width=7.5cm]{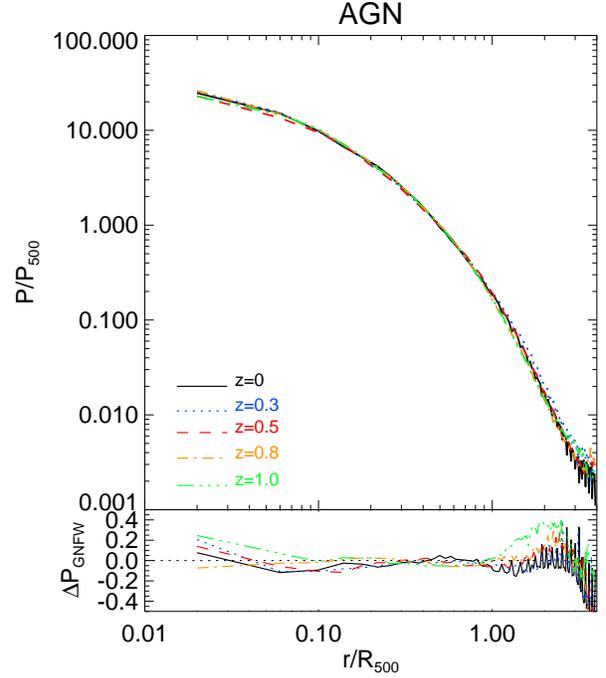}}
\caption{{\it Top panel:} Redshift evolution of the mean pressure
  $P/P_{500}$ radial profiles, out to $4 R_{500}$, of our sample
  of massive clusters in the {\tt \agn} simulations.  Different
  colours are for the results obtained at $z=0, 0.3, 0.5, 0.8$ and
  $1$.  {\it Bottom panel:} Relative difference between the mean
  profiles shown in the top panel at a given redshift and the
  corresponding best--fitting GNFW pressure profile.}
\label{fig:profiles_z}
\end{figure}

As shown by \citet{Nagai_2007_2}, and confirmed by X-ray and SZ
observations of groups and clusters \citep[e.g.][]{Arnaud_2010,
Sun_2011,Planck_2013}, the ICM pressure profiles obtained from
cosmological simulations are well described by a generalized NFW
(GNFW) model:
\begin{equation}
 {p}(x) = \frac{P_{0} } { (c_{500}x)^{\gamma}\left[1+(c_{500}x)^\alpha\right]^{(\beta-\gamma)/\alpha} } \, , 
\label{eq:pgnfw}
\end{equation}
where $x=r/R_{500}$ and the parameters [$P_0$, $c_{500}$, $\gamma$,
$\alpha$, $\beta$] stand, respectively, for the normalization, the
concentration, and the central, intermediate and outer slopes of the
profile.

To analyze the consistency of the mean pressure profiles obtained in
our simulations, we perform a Monte Carlo Markov Chain (MCMC) analysis
to determine the values of the parameters describing the GNFW model of
Eq.~\ref{eq:pgnfw}.  We have performed this analysis for different
sets of fixed and free parameters.  In particular, in order to compare
with different observational samples, we have considered two specific
cases: two four-parameters models where we have fixed in
each case $\gamma$ and $\beta$, respectively.  Therefore, we assume a
first four-parameters model with [$P_0, c_{\rm 500}, \alpha, \beta$]
left free and $\gamma$ fixed to $0.31$, in accordance to the
best-fitting value obtained by \citet{Arnaud_2010}, and a second
four-parameters model described by leaving [$P_0, c_{\rm 500}, \gamma,
\alpha$] free and $\beta$ fixed to $5.0$, which is an intermediate value 
to those obtained by  \citet{Arnaud_2010} and \citet{Planck_2013}.  
We carried out our analysis by considering the
subset of 24 massive simulated clusters, as well as the subsamples of
CC and NCC systems within them. The best fitting parameters we obtain
for each of the considered cases are shown in
Table~\ref{tab:fittings}. In all cases, we have performed the analysis
within the radial range $0.02\mincir r/R_{500}\mincir 4$.

At $z=0$, the results of our fittings are in broad agreement with the
values obtained from previous observational analyses.  In the case
with $\gamma=0.31$, as it is shown in Fig.~\ref{fig:contours_cosmomc},
the best fit values obtained for the {\tt \agn} run show a good
consistency (within 1-$\sigma$) with the best fits obtained by
\citet{Arnaud_2010}, whereas the comparison is slightly worse when looking 
at the \citet{Planck_2013} best-fit values.  In this case, however,
when we include AGN physics, we obtain
for the outer slope $\beta$ a higher value than the one reported by
\citet{Arnaud_2010} \citep[derived from early simulations
by][]{Borgani_2004, Nagai2007, Piffaretti_2008}, whereas our {\tt \nr} 
model produces a better match.
This higher value of the outer slope is in disagreement with the trend
obtained by recent simulations accounting for AGN feedback
\citep[e.g.][]{Greco_2014, LeBrun_2015} or with more recent
\citet{Planck_2013} data, which, as shown in 
Fig.~\ref{fig:contours_cosmomc}, tend to obtain lower values of $\beta$
than reported by \citet{Arnaud_2010}.
Our results on the outer slope are however in better agreement 
with recent observational estimates by \citet{Ramos-Ceja_2015} ({$\beta \sim 6.35$})
or \citet{Sayers_2016} ($\beta \sim 6.13$).  
Given the degeneracies between
the different parameters in this model, it is difficult to perform a
direct and reliable comparison between the best-fitting values
obtained from different studies. However, we note from
Fig.~\ref{fig:contours_cosmomc} that, whenever there is a well defined
degeneracy direction between a pair of parameters, both data and
simulations tend to lie on this direction, with no significant
offset. Therefore, small effects related, e.g. to the radial range
covered by observations and by simulations or residual mass
differences, could cause this tension in the values of the fitting
parameters.

As for the population of CC and NCC clusters within our {\tt \agn}
simulations at $z=0$, the diversity of their corresponding mean
profiles is more evident from the values of the best fit parameters
than from the mean profiles shown in Fig.~\ref{fig:profiles_cc_ncc}.
Indeed, as obtained in previous observational studies
\citep[e.g.][]{Planck_2013}, the mean profiles derived for CC/NCC
systems show the largest discrepancy in the normalization, with CC
having larger values of $P_0$ than NCC clusters.  In general, the
broad agreement obtained between observed and simulated pressure
profiles from the core (with a clear distinction between CC and NCC
systems) out to cluster outskirts, suggests that the thermal structure
of cluster cores is responsible for the scatter and shape of the mean
profiles in central regions, while additional factors, such as the
overall dynamical state, determine the behaviour in the cluster
outskirts. The poor observational characterization of the peripheral
regions and the increasing importance of the non-thermal pressure
support, call for the need of a deeper understanding of the physics
and distribution of the gas in these regions \citep[e.g.][]{Avestruz_2014,Avestruz_2016}, 
both from observations and from simulations.

For completeness, we show in Fig.~\ref{fig:profiles_z} the mean pressure
profiles obtained for our sample of massive clusters in the {\tt \agn}
simulations at $z=0.0, 0.3, 0.5, 0.8,1$.  
The bottom panel of this figure shows $\Delta P_{GNFW}$, that is the relative difference
between the mean profile obtained at each redshift and the
corresponding best-fit to the GNFW model when we consider $\beta=5$.
In general, our best-fit models recover quite well the mean profiles at different redshifts.
For the AGN simulations, the scatter between the recovered and the
input profiles is quite small ($\mincir 5$ per cent) at all redshifts for
$0.03 R_{500} \mincir r \mincir R_{500}$, while it increases
considerably in outer regions and especially at higher redshift.  A
similar result is also obtained for the sample of clusters in the {\tt
\nr} and {\tt \w} simulations.
If we compare our fitting parameters to the results
obtained by \citet{McDonald_2014} for a sample of clusters at $z<0.6$
and $z>0.6$, we obtain a broad agreement, especially at lower
redshifts. However, given the obvious discrepancies between the
different samples and approaches employed, we have to take these
results with caution.  In any case, our results at different redshifts
confirm that self-similar scaling is obeyed by pressure profiles
of massive clusters to a good level of precision.
It is also interesting to note that, as the evolution proceeds, the mean 
pressure profile of our massive systems become slightly steeper, resembling
more the CC associated profiles. This result is consistent with the fact that,
despite the presence of AGN feedback, radiative cooling can not be completely 
halted.     

The deviations shown in the bottom panel of Fig.~\ref{fig:profiles_z}
are partially contributed by the presence of nearby clusters, groups,
substructures or a clumpy gas distribution.  It is also likely that
part of the deviation stems from possible dependencies (which have not
been taken into account) on redshift or mass of the parameters
describing the model.   
However, to analyze properly these
additional dependencies, it is necessary to have a sample of objects
within a larger range of masses and redshifts than the one considered
here.  Deviations from the GNFW model or additional dependencies of the
associated parameters \citep[see e.g.][]{Battaglia_2012b, LeBrun_2015}
would imply a departure from the assumed self-similar evolution
\citep[see also][]{Ramos-Ceja_2015}.

\section{Gas clumping}
\label{sec:clumping}

The existence of small dense clumps and density fluctuations
throughout the ICM can bias high the derivation of the gas density
profile from X-ray surface-brightness observations 
and, as a consequence, the estimation of all the X-ray derived quantities,
such as entropy,  gas mass and pressure
\citep[e.g.][]{Eckert_2013_2}. In a similar way, also the ICM pressure
can deviate from a smooth distribution, affecting the thermal SZ effect signal and 
its scaling relations with cluster mass and the SZ power spectrum
\citep[e.g.][]{Battaglia_2014}. The level of gas density or pressure
inhomogeneity in the ICM is usually characterized by a clumpiness
parameter \citep{Mathiesen_1999}, which, however, is not a directly
observable quantity. In this respect, numerical simulations are useful
instruments to quantify the degree of expected clumpiness, and to investigate its origin
and its impacts on observational measurements  \citep[e.g.][]{Nagai_Lau_2011, Roncarelli_2013, 
Vazza_2013, Zhuravleva_2013, Battaglia_2014}.

\subsection{Definitions}
The clumpiness or clumping factors in gas density, $C_{\rho}$, and
thermal pressure, $C_P$, are usually defined as
\begin{flalign}
  C_{\rho}(r)\,\equiv \frac{\langle \rho^2 \rangle}{\langle \rho \rangle^2} =\,{\sum\limits_i^{N} m_i \rho_i \over \left(\sum\limits_i^{N} m_i \right)^2} V_{shell} 
  \label{eq:clumpiness_rho}\\
  C_{P}(r)\, \equiv \frac{\langle P^2 \rangle}{\langle P \rangle^2}=\,{\sum\limits_i^{N} m_i \rho_i T_{i}^{2} \over \left(\sum\limits_i^{N} m_i T_i \right)^2} V_{shell} \, ,
  \label{eq:clumpiness_p}
\end{flalign}
where the summation is over all the $N$ gas particles within a given
radial shell with volume $V_{shell}=\frac{4}{3} \pi [(r+\Delta r)^3-r^3]$,
and where $m_i$, $\rho_i$ and $T_i$ are, respectively, the mass, density and temperature of
the $i$-th fluid element.  For each halo, we compute the radial profiles 
in 100 equi-spaced bins within the radial range $0.1\le
r/R_{vir} \le 2$.  We have checked that these profiles are almost unaffected when 
we use instead 50 equi-spaced bins. 
By definition, $C_{\rho} \ge 1$ and $C_{P} \ge 1$,
with the case $C_{\rho} =C_{P} =1$ representing a uniform medium. 
Since the distribution of the clumping factor is not Gaussian, in the 
following we will use the median profiles instead of the mean.

In the computation of the gas and pressure clumping factors we only 
take into account the X-ray emitting SPH particles, i.e., those particles 
with $T\ge 10^6 \, K$.  This temperature cut does not significantly affect
the pressure clumping factor obtained from any of our simulations 
\citep[see also][]{Battaglia_2014}. 
 
\begin{figure*}
{\includegraphics[width=15.0cm]{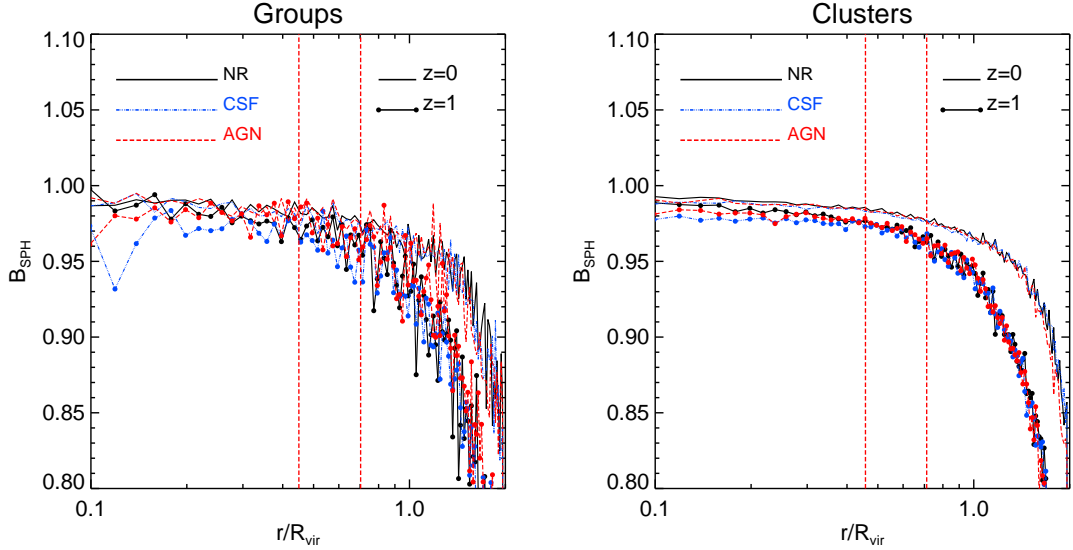}}
\caption{Median radial profiles of the SPH volume bias, as defined by
  Eq.~\ref{eq:sphbias}, out to $2R_{vir}$ for the sample of 5 central
  groups (left panel) and 24 central clusters (right panel) within our
  three sets of simulations. Lines in black, blue and red stand for
  the results at $z=0$ for the {\tt \nr}, {\tt \w} and {\tt \agn}
  simulations, respectively, whereas the same line types connected by
  small dots stand for the results obtained at $z=1$.  Vertical lines
  in both panels represent the mean values of $R_{500}$ and $R_{200}$
  for the {\tt \agn} simulations.}
\label{fig:bias}
\end{figure*}

Recently, \citet{Battaglia_2014} showed that the limitation of
Lagrangian schemes to properly describe an inhomogeneous medium, and
in particular low density regions, generates an incorrect estimation
of the SPH volume, thus affecting the estimation of the clumping
factors. To alleviate this tension, they introduced an SPH volume bias
which is defined as:
 \begin{equation}
 B_{SPH} (r)=\frac{1}{V_{shell}} {\sum\limits_i^{N} \frac{m_i}{\rho_i}}= \frac{1}{V_{shell}} {\sum\limits_i^{N} V_i}\,.
 \label{eq:sphbias}
\end{equation} 
Although this factor only entails a small correction on the SPH volume
calculation within $R_{200}$, it can be more important in outer
regions (Fig.~\ref{fig:bias}). Indeed, both at $z=0$ and $z=1$, the correction
amounts of few percent ($\leq 5\%$) within the virial radius, independently on the mass and ICM physics,
but it rapidly grows to 10\% and 20\% at $1.5$ and 2 times that distance. In these most external regions,
the correction is 5-10\% more significant at $z=1$ than at $z=0$. 
In light of these results, in the following we will correct
the clumping factors given in Eqs.~\ref{eq:clumpiness_rho} and
\ref{eq:clumpiness_p} by the corresponding volume bias:
\begin{align}
C_{\rho}(r)\,= C_{\rho}^{'}(r) \cdot B_{SPH}(r) 
\label{eq:clumpiness_rho_new}\\
C_{P}(r)\,= C_{P}^{'}(r) \cdot B_{SPH}(r)
\label{eq:clumpiness_p_new}
\end{align} 
\noindent
Independently of the inclusion of this correction, as we will see 
throughout this section, our results on clumping are in broad agreement 
with the findings of previous studies performed with either SPH or 
Eulerian-based simulations including different sets of physical processes 
\citep[e.g.][]{Nagai_Lau_2011, Roncarelli_2013, Vazza_2013, Battaglia_2014, 
Eckert_2013_2}.

\begin{figure*}
\centerline{\includegraphics[width=12.0cm]{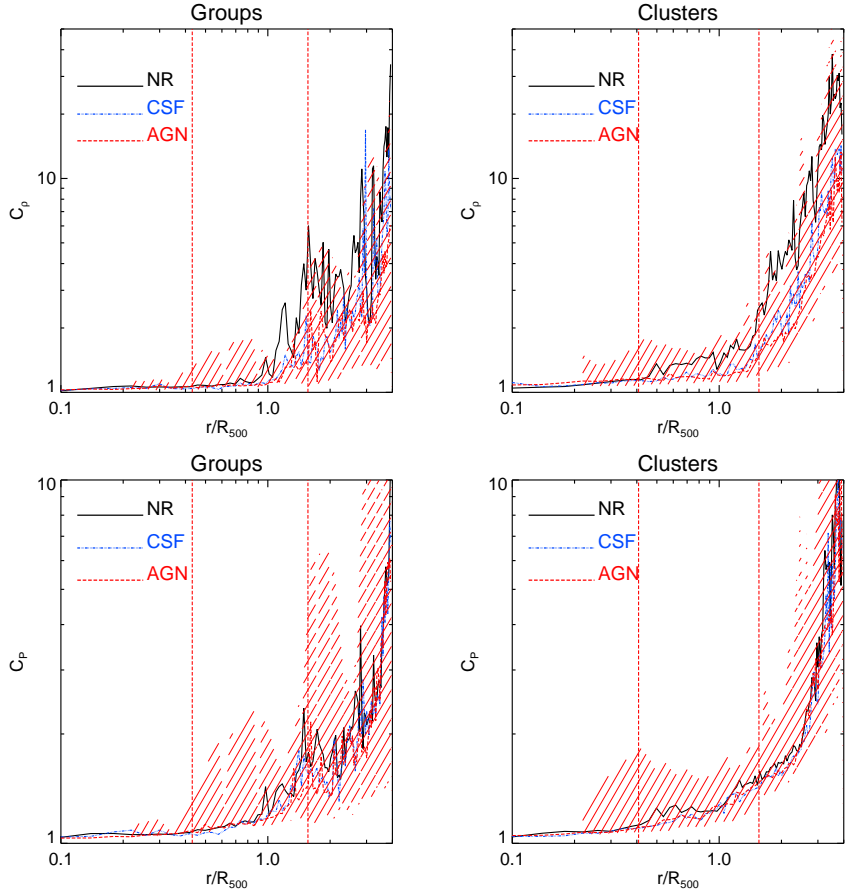}}
\caption{{\it Left column:} Top and bottom panels show, respectively,
  the median radial distribution at $z=0$ of the gas density and
  pressure clumping factors, that is, $C_{\rho}$ and $C_P$ for the
  sample of groups within our simulations.  Black, blue and red lines
  stand for the results obtained for the {\tt \nr}, {\tt \w} and {\tt
    \agn} simulations, respectively.  The coloured area in red stand
  for 1-$\sigma$ dispersion around the median profile of the {\tt
    \agn} run.  Vertical lines represent the mean values of $R_{2500}$
  and $R_{200}$ in units of $R_{500}$ for the sample of considered
  objects within the {\tt \agn} simulations.  {\it Right column:} Same
  results shown in the left column but for the sample of massive
  galaxy clusters.}
\label{fig:clumping_z0}
\end{figure*}

\begin{figure}
\centerline{\includegraphics[width=9.0cm]{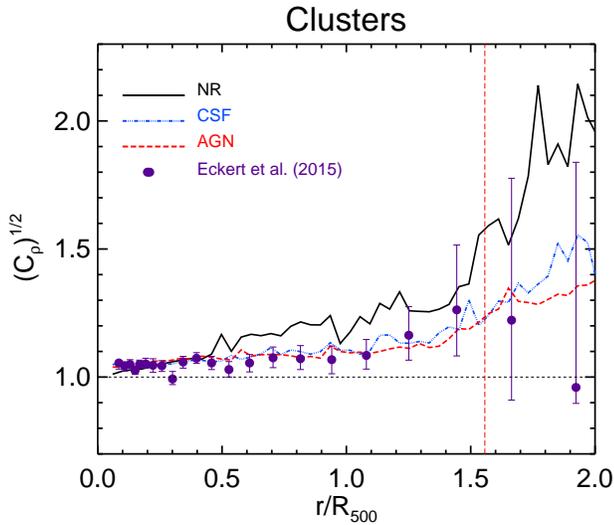}}
\caption{Comparison of the 3D median radial distribution at $z=0$ of
  the gas density clumping factor for the sample of massive clusters
  within our simulations and the observations by
  \citet{Eckert_2013_2}.  Black, blue and red lines stand for the
  results obtained for the {\tt \nr}, {\tt \w} and {\tt \agn}
  simulations, respectively, whereas filled dots represent the
  observational data.  The vertical dashed line represents the mean
  value of $R_{200}$ in units of $R_{500}$ for the sample of
  considered objects within the {\tt \agn} simulations.}
\label{fig:clumping_z0_observed}
\end{figure}

\begin{figure*}
{\includegraphics[width=8.5cm]{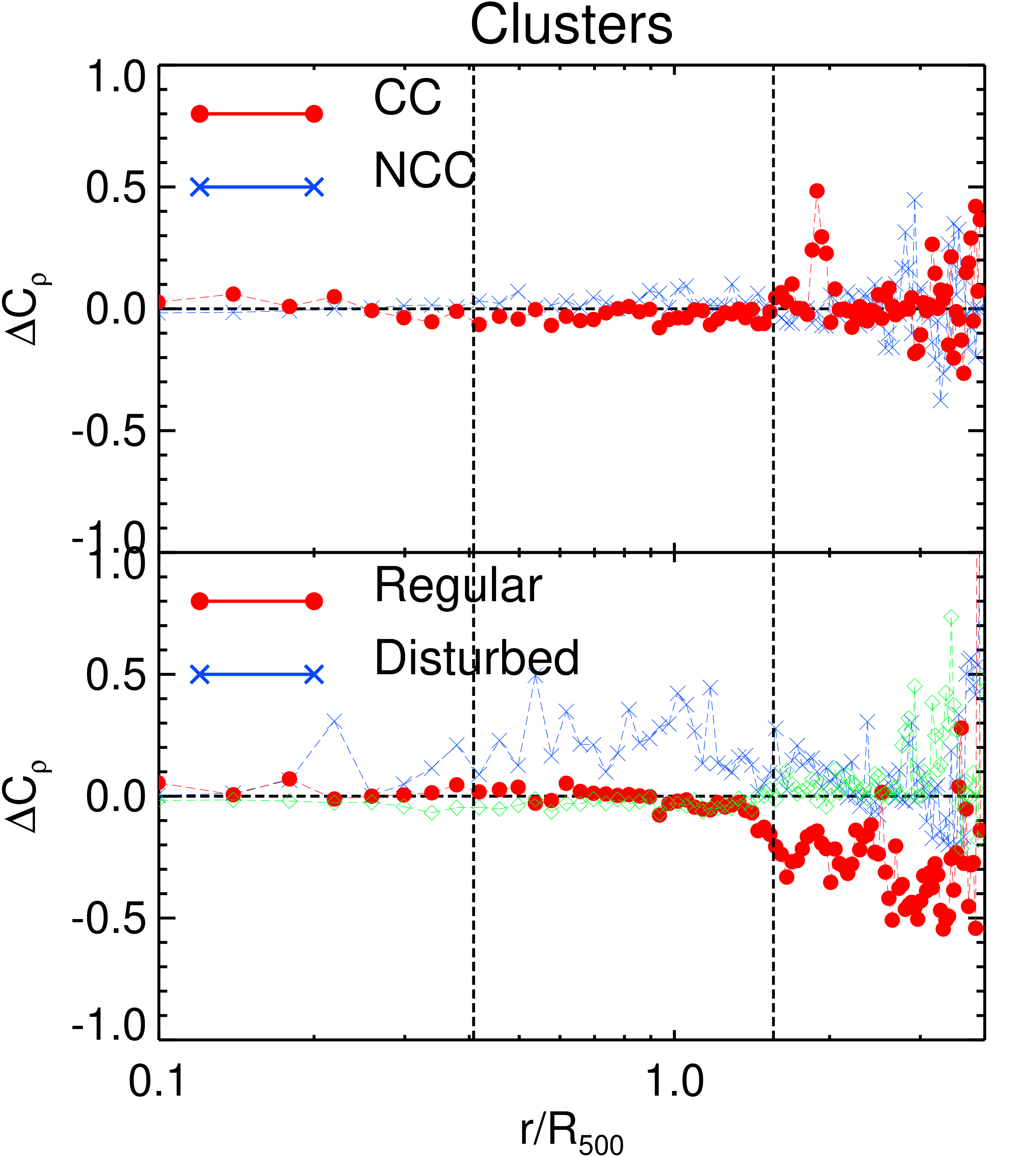}}
{\includegraphics[width=8.5cm]{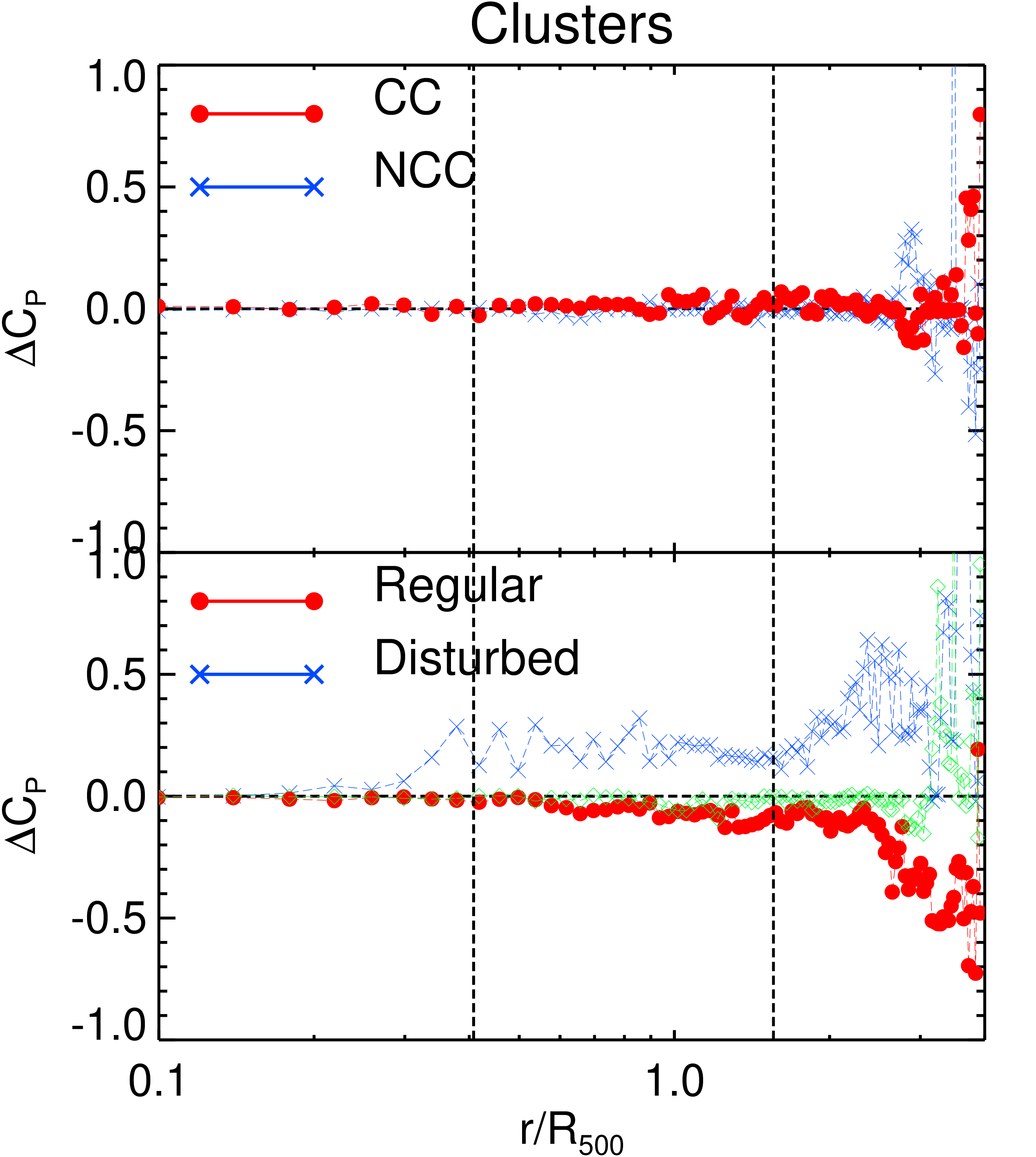}}
\caption{{\it Left panel:} Relative deviation, $\Delta C_{\rho}$,
  between the median profile of gas density clumping obtained for the
  subsamples of CC/NCC and regular/disturbed clusters (top and bottom
  panels, respectively) in the {\tt \agn} simulation with respect to
  the corresponding global median profile shown in
  Fig.~\ref{fig:clumping_z0}. Profiles for CC and relaxed systems are
  represented in red by filled circles, whereas NCC or disturbed
  clusters are given by blue lines with crosses. Green lines in the
  bottom panel stand for the sample of clusters classified as in an
  intermediate state.  {\it Right panel:} The same as in the left
  panel but for the corresponding deviation in pressure clumping,
  $\Delta C_P$. Vertical lines indicate the mean values of $R_{2500}$
  and $R_{200}$ for the sample of considered clusters.}
\label{fig:clumping_dynamical}
\end{figure*}

\subsection{Results on clumping}

Figure \ref{fig:clumping_z0} shows the radial distribution of $C_{\rho}$ and 
$C_P$ obtained for the sample of 5 central groups and 24 central clusters
within our three sets of simulations at $z=0$.   
At the mass scale of groups, the gas density clumping factor shows
values smaller than $\sim 2$ within $R_{200}$.
Generically, the two radiative simulations produce similar results, with systematically
lower degree of clumping than in the non-radiative case. In fact,
radiative cooling has the effect of removing high--density, relatively
cold gas from the hot phase, thus suppressing gas clumping. At the
same time, feedback from SN driven winds and, even more, from AGN
displace gas from high--density regions, thus further contributing to
smooth the density and thermal distributions of the ICM \citep[see
e.g.][]{Planelles_2013_b, Rasia_2014}. As a consequence, there is a
reduction in the number of small cold density clumps which are more
prominent in outer cluster regions. 
These dense clumps are mainly
associated with substructures being accreted to the centre of more
massive central systems.  This is the reason why the clumpiness is
more important in outer regions, which are dynamically younger and
still under the effects of ongoing matter accretion, than in inner
regions. 
The density clumping factor shows a
similar behavior in clusters and in groups within $R_{200}$. However,
beyond this radius, $C_{\rho}$ in clusters is higher and steeper than
in smaller systems.  This is due to the fact that massive systems are
more prone to significant merging and matter accretion since they are,
on average, formed more recently.

\vspace{0.5cm}

Measurements of  density clumping are
quite uncertain. Indeed, only the application of indirect methods has led recently
to some estimations.  \cite{Simionescu2011} analyzed {\it Suzaku}
observations of the outskirts of the Perseus cluster. They measured
values of gas fraction in the outer regions well in excess of the
cosmic value, a result that they interpreted as spuriously induced by
gas density clumping, with a clumping factor of $C_{\rho}\sim 1-3$ at
$R_{500}$ and $C_{\rho}\sim 9-12$ at $1.5\times
R_{500}$. \citet{Walker_2012} also obtained relatively large values
for the clumping, with $C_{\rho}\sim 1-3$ at $R_{500}$ and
$C_{\rho}\sim 2-9$ at $1.5\times R_{500}$ for the PKS0745-191 cluster
\citep[cf. also][]{Morandi_2013}.  On the contrary,
\cite{Eckert_2013_1} used the deviations from self-similarity of the
observed entropy profiles for a sample of 18 clusters for which both
\rosat\ measurements of X--ray surface brightness and \planck\
measurements of the SZ signal are available beyond $R_{500}$. 
From their analysis, \cite{Eckert_2013_1} inferred values of
clumping $C_{\rho}\sim 1.2$ at $R_{200}$, thus considerably lower than
found by \cite{Simionescu2011}. In a further analysis,
\cite{Eckert_2013_2} used a slightly larger sample of clusters
observed with \rosat\, and \planck\, together with a new technique to
obtain unbiased density profiles from X-ray data. Their newly derived values, 
$\sqrt{C_{\rho}}< 1.1$ for $r<R_{500}$, are thus in good agreement with
their previous measurements implying a modest gas density clumping.

It is with this large statistical sample that we compare our simulations in
Fig.~\ref{fig:clumping_z0_observed}. 
Since the dispersion around the median clumping profiles is significant 
for our three models, we should take these results with caution. However, 
by looking at the median clumping profiles obtained for each set of simulations, there is a clear trend:
both radiative simulations are in remarkably good agreement with the data
whereas the {\tt \nr} runs produce systematically larger values ($\magcir 20\%$ beyond $R_{200}$) than
observed \citep[see also][]{Vazza_2013}. In general, results on
density clumping from our radiative simulations favour small values of
gas clumping, independently of the feedback scheme included. 
Thus, they are at variance with the large clumping factors
found by \cite{Simionescu2011} from {\it Suzaku} observations.

The bottom panels of Fig.~\ref{fig:clumping_z0} show that the pressure clumping behaves similarly to
$C_{\rho}$, with the exception that the radial
behavior of $C_P$ does not depend significantly on the physical
model.  This suggests that the small cold density clumps that enlarge the density
clumping in outer regions of clusters and groups within
our {\tt \nr} simulations are in pressure equilibrium with the
surrounding medium and, thus, give a negligible contribution to the
pressure clumping \citep[see also][]{Battaglia_2014}.  In addition, although $C_P$ is slightly larger
than $C_{\rho}$ within $R_{200}$, it shows lower values outside this region.

\vspace{0.3cm}

Given the obvious connection between clumpiness and environment, it is
interesting to analyze how it relates to the thermal or dynamical
state of each cluster.  To this purpose, we now analyze the
distribution of  clumpiness in the {\tt \agn} simulations when
clusters are divided in CC/NCC and in regular/disturbed objects (see
Section~\ref{subsec:sample}). For each of these subsamples,  we
have computed the radial distributions
of gas density and pressure clumpiness, that is, $C_{\rho_{sub}}$ and
$C_{P_{sub}}$. We compare these median profiles to the corresponding
global median profile shown in Fig.~\ref{fig:clumping_z0}.  Figure
\ref{fig:clumping_dynamical} shows the relative difference between
the median profiles in gas density clumpiness (left panels), $\Delta
C_{\rho}=(C_{\rho_{sub}}- C_{\rho})/C_{\rho}$, and pressure clumpiness
(right panels), $\Delta C_P=(C_{P_{sub}}-C_P)/ C_P$, for the sample of
CC/NCC (upper plots) and regular/disturbed systems (lower plots).
As for the gas density clumpiness, when our sample is divided into CC
and NCC-like clusters, there is no much difference between these two
populations, showing the larger differences at $r\geq
R_{200}$. Moreover, the two subsamples show negligible differences in
clumpiness throughout the whole radial range, indicating that the
radial distribution of $C_{\rho}$ in both cases is quite similar to
the distribution of clumpiness obtained for the whole sample of
clusters.  Only at $r\geq R_{200}$, $\Delta C_{\rho}$ slightly
deviates from zero, especially for the CC systems.
On the contrary, if we analyze the radial distribution of $\Delta
C_{\rho}$ obtained for the populations of regular and disturbed
clusters, there is a clear difference between them. In particular,
while regular objects show $\Delta C_{\rho} \sim 0$ out to $\sim
R_{200}$, disturbed clusters show a larger deviation already from
$\sim 0.2 R_{500}$. Whereas
regular systems always tend to produce $\Delta C_{\rho} \leq 0$,
disturbed clusters tend to show larger values of clumpiness and a
larger cluster-to-cluster scatter than the global sample.
A similar behaviour is obtained for $\Delta C_P$. In this case,
however, both CC and NCC systems show $\Delta C_P \sim 0$ at least out
to $\sim 3R_{500}$, whereas the deviation between regular and
disturbed clusters is already evident from $\sim 0.3R_{500}$.

These results indicate that, whereas density and pressure clumpiness
is quite insensitive to the thermal properties of clusters in their
core regions, it depends significantly on their global dynamical
state. As shown in \citet{Biffi_2016}, this trend is also consistent with the
HE deviation obtained for the same sample of clusters.
In general, dynamically disturbed objects are still being formed and,
therefore, their ICM is more inhomogeneous than in dynamically relaxed
systems.  Indeed, in previous studies the degree of ICM density
clumping has been used as a criterion to distinguish between relaxed
and unrelaxed systems \citep[see][]{Roncarelli_2013}.
In this regard, \citet{Battaglia_2014} also found a similar  dependence on 
the dynamical state of their clusters, with relaxed systems showing the lower clumping values.

\begin{figure*}
{\includegraphics[width=8.5cm]{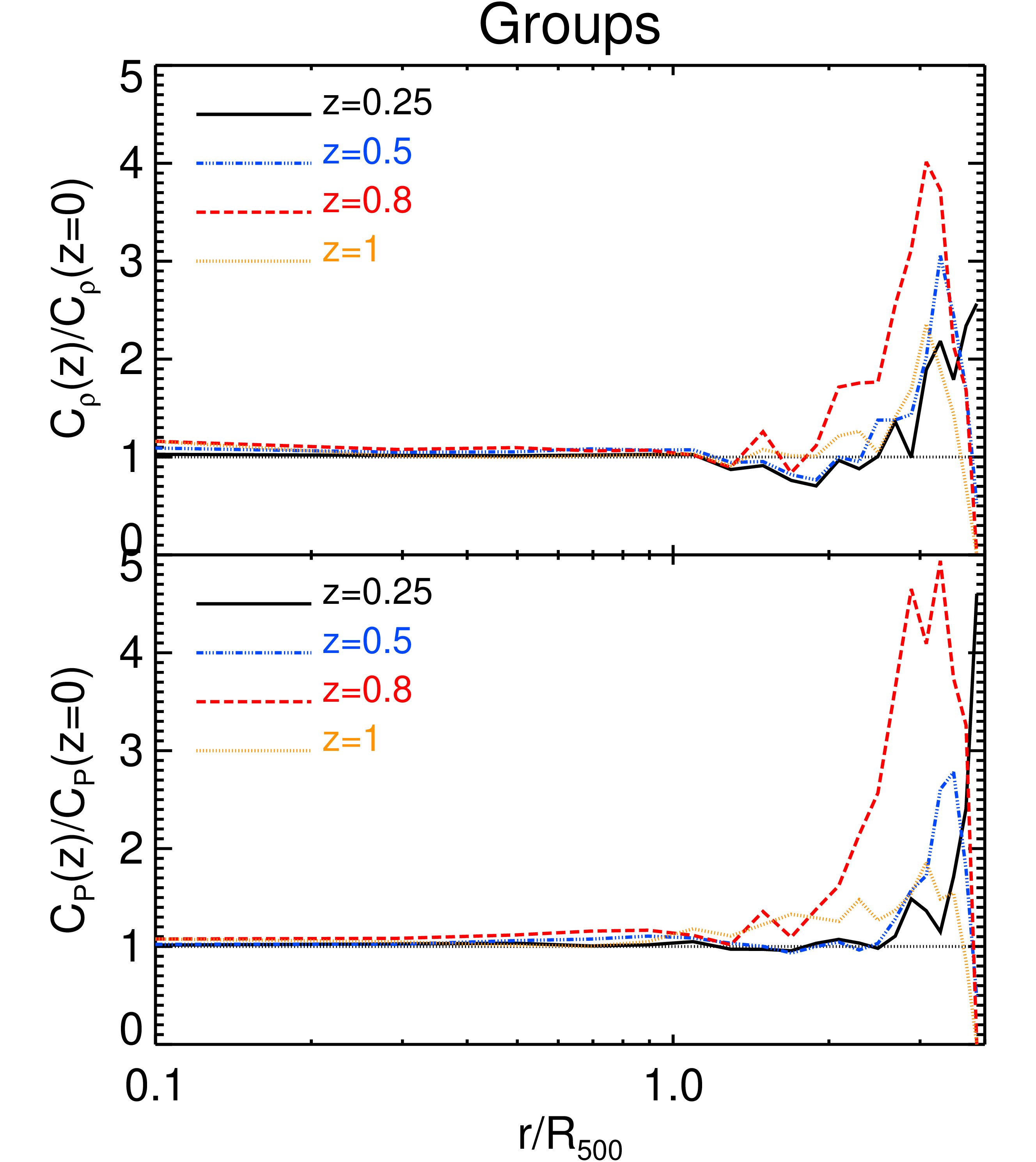}}
{\includegraphics[width=8.5cm]{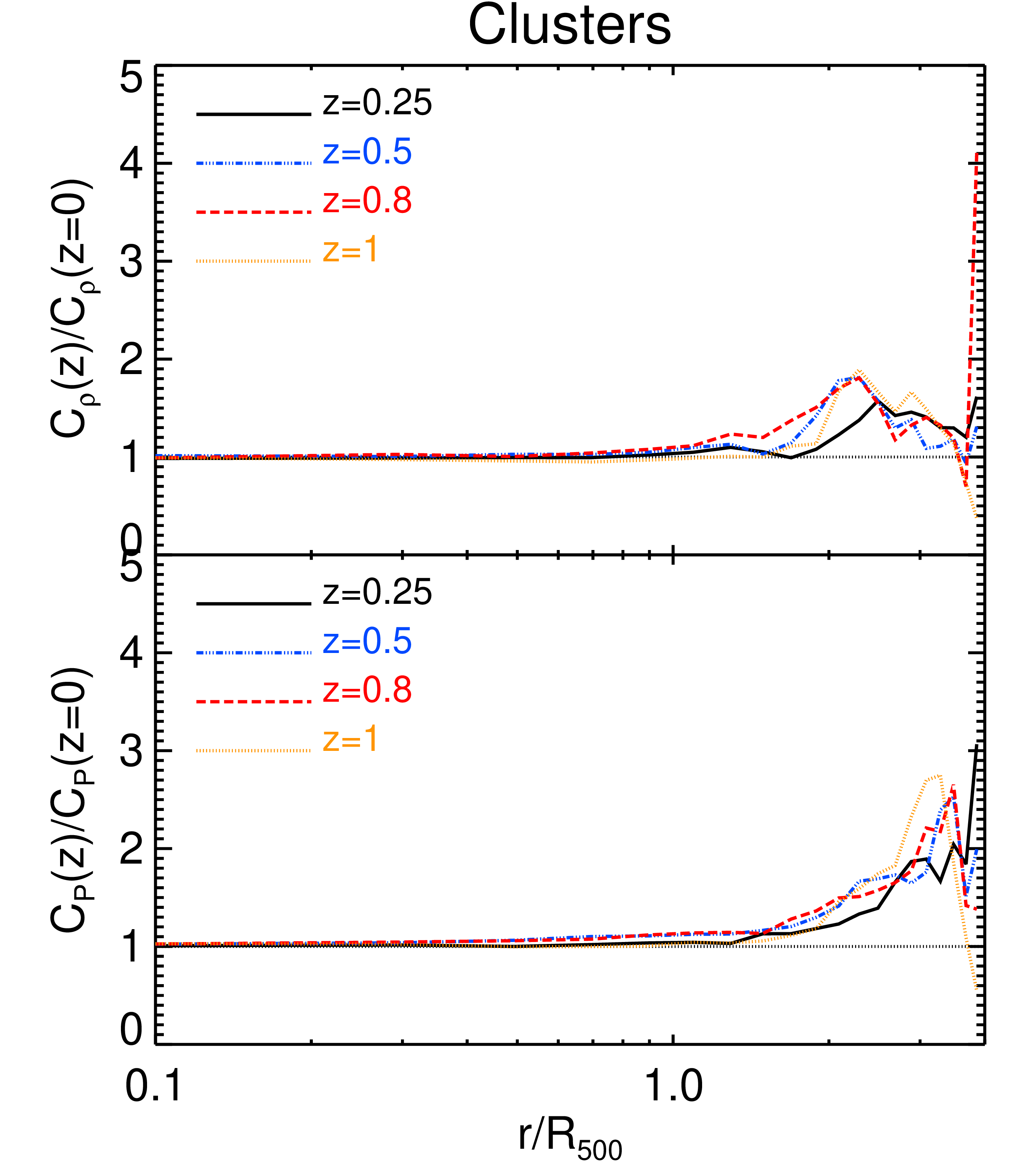}}
\caption{{\it Left column:} Top and bottom panels show, respectively,
  the ratio between the radial distribution of the gas density and
  pressure clumping factors at different redshifts, and the
  corresponding values at $z=0$ for the sample of groups within our
  {\tt \agn} simulations.  Different line types stand for the ratios
  obtained at $z=0.3, 0.5, 0.8$ and 1.  {\it Right column:} The same
  as in the left column but for the sample of massive galaxy
  clusters.}
\label{fig:clumping_z_evol}
\end{figure*}

Figure ~\ref{fig:clumping_z_evol}  shows that the density
clumping factor generally increases with redshift, especially at $r
\ge 2\times R_{500}$.  This increase, which is slightly larger for our sample
of groups, amounts to a factor of $\sim4$ at $z=0.8$.  A similar trend
is also obtained for $C_P(r)$ for both groups and clusters.
Given the relatively low number of systems and the large scatter of the profiles,
our results have to be taken with caution.
Nevertheless, a more pronounced clumping at high redshift, both in density and
pressure, is in line with the younger age of systems at earlier
epochs, when accretion from substructures and filaments is more
efficient \citep[see also][]{Battaglia_2014}. 
A stronger increase of clumping for groups is consistent
with the expectation that ram-pressure stripping is less efficient
there than in clusters, thus causing a less efficient removal of gas
from merging clumps.

Therefore, as predicted by simulations and confirmed by recent
observational determinations, there is a certain level of clumpiness in 
the ICM that should be considered for precise X-ray and SZ measurements. 
In our simulations its degree is considerably lower than some values advanced 
to explain {\it Suzaku} observations and it is still consistent with other X-ray and SZ 
measurements of a local sample of clusters.
However, caution in interpreting projected quantities should be applied as the 
clumpiness grows very rapidly outside the virial radius.

\section{SZ scaling relations}
\label{sec:SZ}

%
After analyzing the pressure profiles and the gas clumping in our
simulations, we consider now how these results translate in the SZ
scaling relations.  As anticipated in Section \ref{subsec:sample}, to
increase the statistics and to enlarge the mass range, here we will utilize 
the complete sample of simulated groups and clusters.

\subsection{Definitions}
\label{sec:sz_def}

From the combination of X-ray and SZ data it is possible to derive
the deprojected profiles (e.g. for pressure and temperature).
Usually, in order to do that, models based on the
$\beta$-model \citep[e.g.][]{Bonamente_2008} or on more accurate
modelling of the gas are used \citep[e.g.][]{Mroczkowski_2009}.
In our simulations we compute the cylindrical $Y$ (from now on simply referred to
as \ysz), which is the integral of the Compton-y parameter over
the angular cluster extent, and is defined as the integral of 
the product of
the electron density and the electron temperature profiles over
the considered volume, i.e.,
 \begin{equation}
 Y_{SZ} =  \int y \, d\Omega \,=  \frac{1}{D_A^2}\frac{\sigma_T k_B}{m_e c^2} \int n_e(r) T_e dV \,,
 \label{eq:Ysz}
\end{equation}
where $\sigma_T$, $m_e$, $k_B$, $c$ and $D_A$ are, respectively, the
Thomson cross-section, the electron mass, the Boltzmann constant, the
speed of light, and the angular distance of the cluster.

We evaluated \ysz\ from Compton-y 2-D maps obtained with 
a map making utility for idealized observations detailedly described in
\cite{Dolag_2005}. The Comptonisation parameter maps were computed using all gas particles
in a cuboid of $5R_{500} \times 5R_{500} \times 10R_{500}$ (the last dimension is 
along the line-of-sight).
Each projection was centered on the cluster center, namely
the location of the particle with the minimum gravitational potential as in the
previous analysis.
We calculate projected Compton-y maps along the three main axes for each cluster at 
redshift $0, 0.25, 0.5, 0.8$ and $1$. 
$Y_{\rm SZ,500}$ and $Y_{\rm SZ,2500}$ were calculated from 2-D maps by integrating the Compton-y 
in circles with radii $R_{500}$ and $R_{2500}$.

The thermal SZ flux measures the global thermal pressure of the hot
intra-cluster plasma, representing a close proxy
for the cluster gravitational potential and, therefore, for the
cluster total mass. As derived from Eq.~\ref{eq:Ysz}, the integrated
thermal SZ is proportional to the product between the gas mass (over
the selected integration volume) and the electron temperature 
(corresponding in simulations to the mass-weighted temperature):
$Y_{SZ} D_A^2 \propto M_{gas} T_{mw} $.  If we assume HE
and isothermal temperature distribution, we can use the
scaling between temperature and total mass, $T_{mw} \propto M_{tot}^{2/3}
E(z)^{2/3}$, to finally obtain
\be
Y_{SZ} D_A^2 \propto M_{gas} T_{mw}  \propto M_{gas} M_{tot}^{2/3} E(z)^{2/3} \, ,
\ee
where, as usual, $E(z)$ describes the evolution of the Hubble parameter. 
Since $M_{gas} \propto M_{tot}$ in the self-similar scenario, we expect the slope of 
the \ysz$-M$ relation to be $5/3$ at a fixed redshift.

In the following, we will report the results of our simulations using
the cylindrically integrated \ysz. In particular, in the next two
Sections we will focus on the relation between \ysz\ and the
cluster total mass and its evolution, whereas in Section
\ref{sz_clumps} we will show the relation between \ysz\ and its
X-ray analogue, i.e., $Y_{X} = M_{gas} T_X$, where for $T_X$ we used the
core-excised spectroscopic-like temperature $T_{sl}$ evaluated over the radial
range $0.15-1 \; R_{500}$.
In each case, the best-fit relations are obtained by fitting a
power-law of the form: 
\be E(z)^{-2/3} \, Y \, = \, 10^A \,
\left(\frac{X}{X_0}\right)^B \, ,
\label{eq:power_law}
\ee
where $X$ is the selected quantity to be related with Y, $X_0$ is the
pivot for the same quantity, and $A$ and $B$ describe the
normalization and the slope of the scaling relation, respectively. The
scatter on the best-fit relation between the SZ flux and the cluster
property $X$ is obtained as
\be
\sigma_{\log Y} = \sqrt{\frac{\sum_i^N (\log(Y_i) - (A \cdot \log(X_i) + B ))^2}{N-2}}
\label{eq:sigma_power_law}
\ee
where the index $i$ is running over all the clusters.

\subsection{The \ysz$-M$ scaling relation}
\label{sec:sz_scaling}

In the last decade a number of analyses of SZ observational data have
been carried out to determine the scaling relation between the SZ
signal and the total mass of clusters \citep[see e.g.][for a
review]{Giodini_2013}. The main outcomes of these analyses are that the
SZ effect correlates strongly with mass and that the evolution of the
$Y_{SZ}-M$ relation agrees with the prediction from self-similarity.
The intrinsic scatter is found to vary between $12$ per cent
\citep{Hoekstra_2012} and $20$ per cent \citep{Marrone_2012}, when
cluster lensing masses are used, while in the combined SZ/X-ray study
by \citet{Bonamente_2008} the scatter is even lower, $\sim 10$ per
cent, when evaluated for masses obtained from HE.
Recently, \cite{Czakon_15}, using the Bolocam X-ray-SZ sample, 
reported a $\sim 25\%$ scatter and a discrepancy with the self-similar 
prediction that appears to be caused by a different calibration of 
the X-ray mass proxy with respect to other analyses.
The larger scatter reported by \citet{Marrone_2012} was explained as
partially due to cluster morphology, while most of the segregation is
caused by modeling clusters as spherical objects.  Instead, in a
previous work on a smaller set of $14$ clusters, \citet{Marrone_2009}
found no clear evidence of morphological difference with respect to
cluster dynamical state (classified as disturbed/undisturbed based on
the X-ray-SZ peak offset). 
Simulations also tend to support the self-similar
description for the \ysz$-M$ relation, especially on cluster scales,
where several studies \citep[e.g.][]{daSilva_2004, Nagai_2006,
Battaglia_2012a, Kay_2012, sembolini_etal13, Pike_2014, Hahn_2015}
have found the relation to be robust when non-gravitational physics is
included. However, with respect to observations, simulations predict 
a lower intrinsic scatter with a value that spans from $10$ per
cent \citep{Nagai_2006} to less than $5$ per cent
\citep{sembolini_etal13}.  When doing this comparison, it is worth
keeping in mind that simulation results are based on true masses,
whereas observational scaling relations are computed 
either through X-ray or gravitational lensing.

\begin{figure*}
{\includegraphics[width=9.6cm]{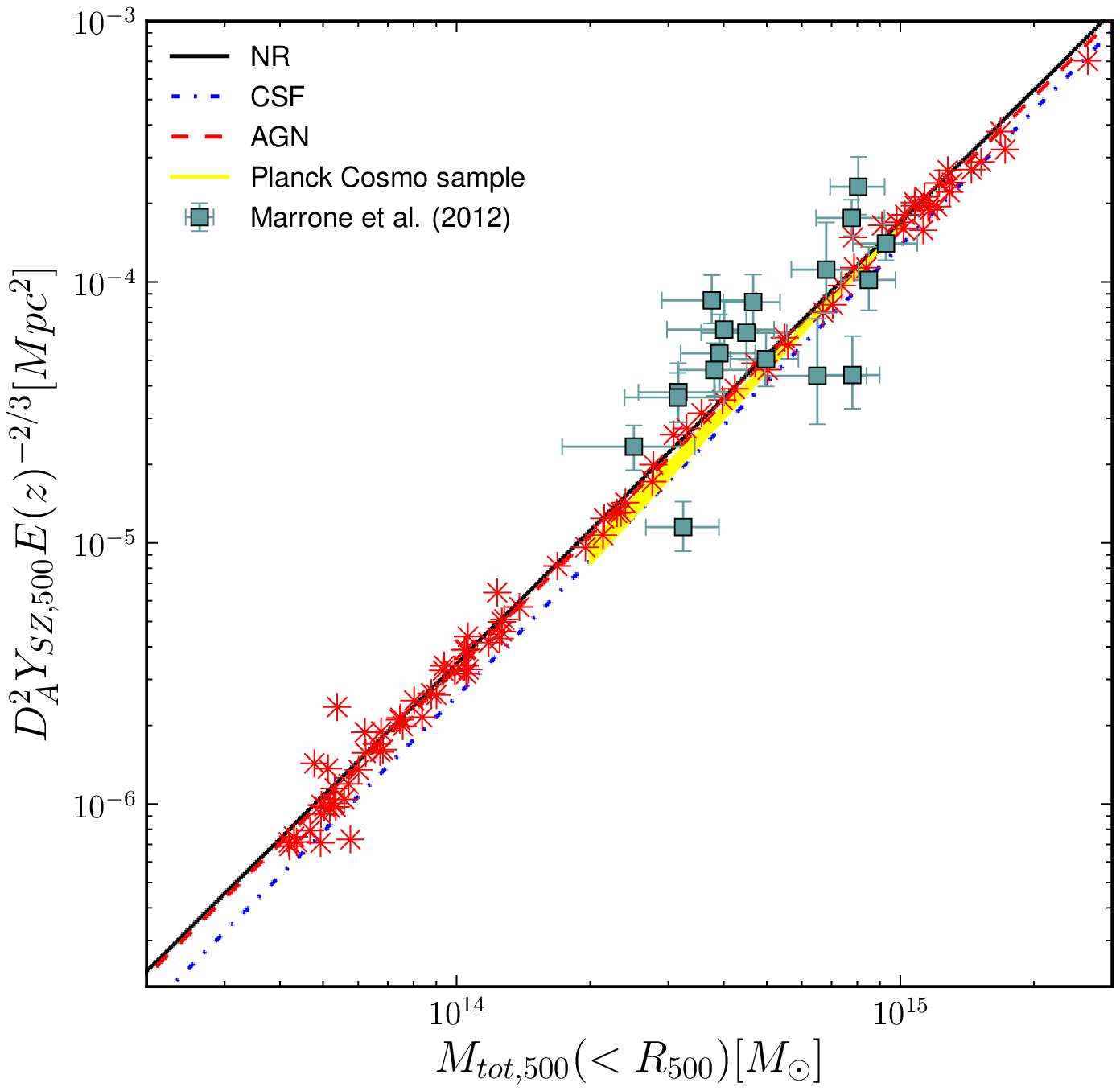}}
\hspace{-1.8cm}
{\includegraphics[width=9.6cm]{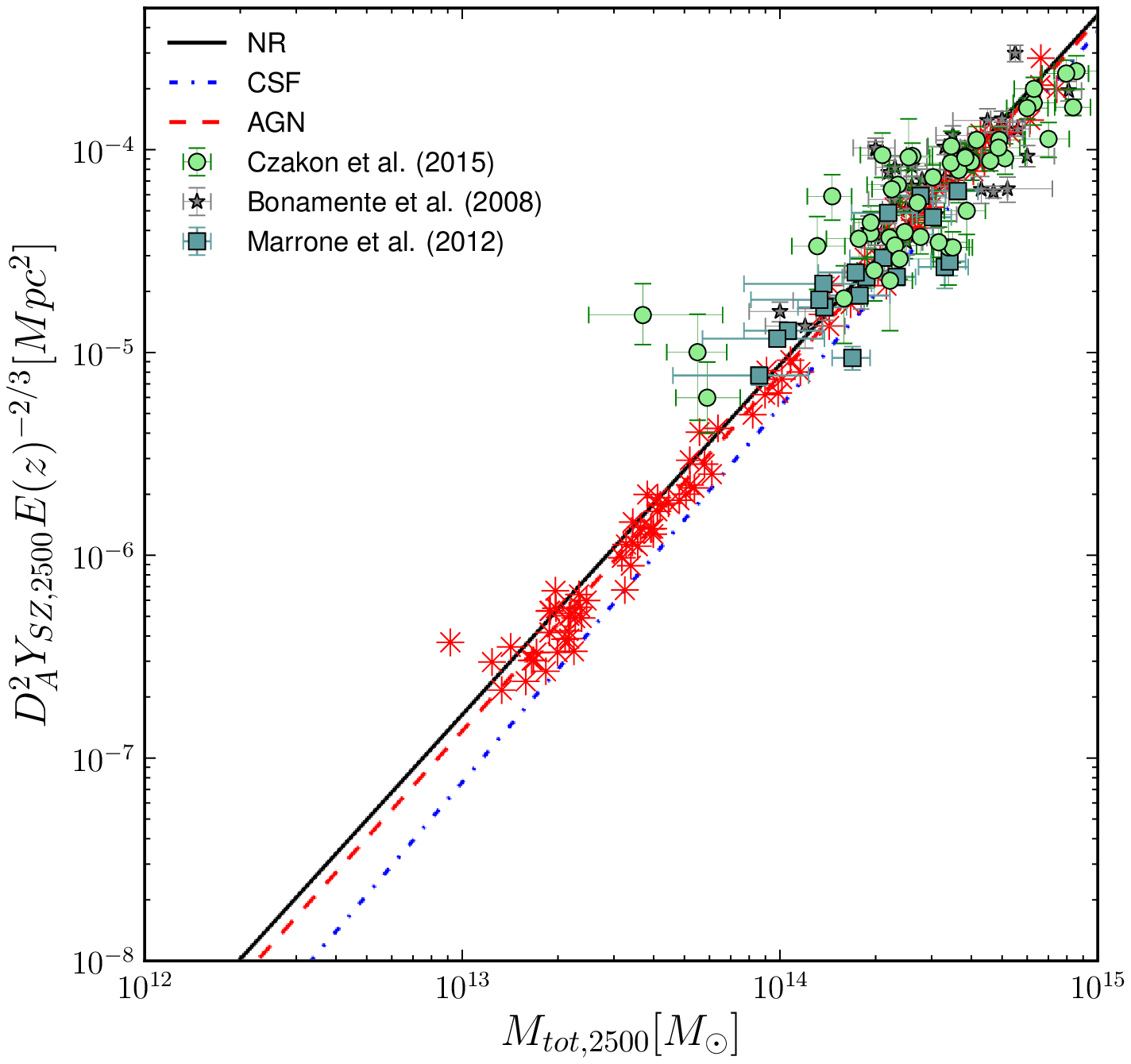}}
\caption{Relation between SZ flux and cluster mass within $R_{500}$
  (left panel) and $R_{2500}$ (right panel) for the sample of
  simulated clusters.  SZ results from simulations refer to the
  cylindrical $Y$. We plot the best-fit relation obtained for the {\tt
    \nr} (black continuous line), {\tt \w} (blue dot-dashed line) and
  {\tt \agn} (red dashed line) simulations. For the sake of clarity,
  only data for the {\tt \agn} run is plotted (red asterisks).
  Different observational samples are used for comparison. In both
  panels, results from the analysis by \citet{Marrone_2012} from
  LoCuSS-SZA clusters are shown as light-blue squares with error bars.
  In the left panel, we also compare our simulations with the best-fit
  relation obtained for the Planck Cosmo sample (yellow shaded area;
  \citealt{Planck_2013_a}).  In the right panel, we show the
  Chandra-BIMA/OVRO cluster sample by \citet{Bonamente_2008} (stars
  with error bars) and the Bolocam X-ray SZ sample BOXSZ by
  \citet{Czakon_15} (green dots with error bars). }
\label{fig:Y-M500}
\end{figure*}

In the left panel of Fig.~\ref{fig:Y-M500} we compare the $Y_{\rm SZ,
  500}-M_{500}$ relation obtained for our simulated cluster set with
observations from \citet{Marrone_2012} and the Cosmo sample from
\citet{Planck_2013_a}.  The best-fit relations obtained in the {\tt
  \nr}, {\tt \w} and {\tt \agn} simulations are shown with different
line types. In Table~\ref{tab:Y-Mfittings} we report the best-fitting
values for normalization, power-law index and scatter obtained in each
case.  For the sake of clarity, we only show the data points for the
{\tt \agn} simulation (red asterisks).  We note that the best-fit
relations obtained for our different models are close to each other
over all the mass range sampled by our simulations. This is in
agreement with the mean weighted pressure profiles shown in
Fig.~\ref{fig:different_masses}, which also demonstrates that the main
contribution to the SZ signal comes from radii around $R_{500}$, consistently 
with \citet{Kay_2012}, \citet{sembolini_etal13} and \citet{Pike_2014}.

In the same figure, we note some differences in the
relation obtained from simulated data and observations. In particular,
the 18 clusters observed by \citet{Marrone_2012} in the redshift
range around $z \sim 0.2$ are plotted as grey squares with error
bars. These results are based on weak-lensing cluster masses
obtained from observations with the \subaru\ telescope and on \ysz\ 
computed from observations with the \sza\ (SZA). 
The simulations agree quite well with the data in normalization and 
slope but not in the scatter which is of order of $20\%$ for 
\citet{Marrone_2012} and $4-5\%$ in our simulated sample (see
Table~\ref{tab:Y-Mfittings}).  

In Fig.~\ref{fig:Y-M500} we also compare simulation results with those
from the Planck Cosmo sample as presented in
\citet{Planck_2013_a}. This sample includes 71 clusters from different
studies: the Early SZ cluster sample observed by Planck
\citep{Planck_XI_2011}, a smaller set of LoCuSS clusters
\citep{Planck_IntIII_2013}, and a sample of \xmm\ clusters
\citep{Planck_IntIV_2013} observed by Planck. The reported best-fit
relation (in yellow) was obtained by applying the correction for the
Malmquist bias. Observations report an intrinsic orthogonal scatter
$\sigma_{\log Y}$ at fixed mass of $6.3$ per cent, thus similar to our
best-fit scatter reported in Table~\ref{tab:Y-Mfittings}. However,
Planck data show a slope, $B=1.79$, which is
steeper than the typical values found in our simulations,
$B=1.68-1.73$.

In the right panel of Fig.~\ref{fig:Y-M500} we report the same
relation but for $\Delta=2500$, i.e, $Y_{\rm SZ, 2500}-M_{2500}$.  In
this case, we compare results from our simulations with three different
observational samples: the previously mentioned LoCuSS/SZA cluster
sample by \citet{Marrone_2012} (light-blue squares with error bars), 
the Chandra-BIMA/OVRO cluster sample analysed by
\citet{Bonamente_2008} (grey stars with error bars) and the 
results regarding the BOXSZ sample reported 
by \citet{Czakon_15} (green dots with error bars).
The latter sample is composed by 45 galaxy clusters  
with a median redshift of $\langle z\rangle \sim 0.42$, that were 
observed with Bolocam and for which \chandra\ data were used. 
\cite{Czakon_15} report a 
5-$\sigma$ shallower slope for their $Y_{\rm SZ, 2500}-M_{2500}$
relation than is predicted by the self-similar model. 
A similar discrepancy was also recently found by 
\cite{Romero_2016}, who used a smaller set of clusters observed
with MUSTANG and Bolocam.

Within these inner cluster regions, we see a distinction between  
models, with a lower normalization in the
case of {\tt \w} clusters that is particularly visible when looking
at systems with $M_{2500}<10^{14} M_{\odot}$.  This trend is in
qualitative agreement with what obtained in the pressure profiles weighted by
radius (right panel of Fig.~\ref{fig:different_masses}), where {\tt
\agn} and {\tt \w} central groups show slightly lower pressure
values (this effect is stronger when we consider the whole sample of
groups and not only the isolated ones). The result on pressure
profiles translates into a lower integrated cylindrical \ysz\ for
galaxy groups. For clusters with larger masses ($M_{2500}> 3 \times 
10^{14} M_{\odot}$) the difference is not so pronounced, and even reduced 
by almost a half in the {\tt \w} case. However, we would like to 
stress again that the SZ properties we present here refer to the `complete' sample, 
while the results presented in Fig.~\ref{fig:different_masses} 
are for the reduced sample of systems.

Here again we note some discrepancy between the observational
datasets. 
The analysis by \citet{Czakon_15} on BOXSZ data 
covers the highest mass range and reports a shallow 
slope ($B_{2500}=1.06\pm0.12$) and a 25\% scatter in data.
As for the analysis by \citet{Bonamente_2008} 
data show a large scatter and the best-fit
relation ($A_{2500}=-3.83$, $B_{2500}=1.66\pm0.20$) 
has a slightly higher normalization and a shallower slope.
\citet{Marrone_2012} instead obtain a slope and normalization
($A_{2500}=-4.56$, $B_{2500}=1.81$) in close agreement with our
results.
It is worth pointing out that masses in the analysis by
\citet{Czakon_15} do rely on the $M_{gas}$ proxy, those from
\citet{Bonamente_2008} are based on X--ray analysis and assumption of
HE, while \citet{Marrone_2012} use masses derived
from weak lensing analyses.  Since we use true masses in our analysis,
we expect better agreement with \citet{Marrone_2012} given that, 
in principle, weak lensing measurements provide mass estimates 
closer to  true cluster masses, provided that systematics are sufficiently under control 
\citep[e.g.][]{Meneghetti_2010, Becker_11, Rasia_2012, Mantz_16}. 
\citet{Czakon_15} report instead that 
most of the discrepancy with respect to the self-similar 
scaling is caused by differences in the calibration of the X-ray mass
proxies adopted in different analyses of observational data.
The tension of simulations with the results by
\citet{Bonamente_2008} could be alleviated by accounting for the
effect of hydrostatic bias, which causes X--ray masses to be
underestimated by 10--20 per cent \citep[e.g.][and references
therein]{Rasia_2012}. 
Accounting for a possible hydrostatic bias in a self-consistent way
would clearly require that also the value of the characteristic radii,
$R_{500}$ and $R_{2500}$, should be decreased accordingly. A proper
inclusion of the effects of hydrostatic bias on the scaling relations
is beyond the scope of this paper and we refer the reader to specific
analyses devoted to it \citep[e.g.][and references
therein]{Biffi_2016}.

\begin{table*}
  \caption{Best-fit parameters for the normalization, $A$, the slope, $B$,  and the scatter, $\sigma_{log_{10}}$, 
    (as introduced in Eqs.~\ref{eq:power_law} and \ref{eq:sigma_power_law})
    describing the relation between \ysz\ and cluster mass evaluated within $R_{500}$ and $R_{2500}$. 
    The pivot $X_0$ is equal to $5\times10^{14} M_{\odot}$ and $2\times10^{14} M_{\odot}$ 
    (for $R_{500}$ and $R_{2500}$, respectively).
    The parameters are obtained  for the complete sample of clusters and groups within 
    the {\tt \nr}, {\tt \w} and {\tt \agn} simulations. For
    completeness, we also show the best-fit parameters obtained for
    the subsamples of regular/disturbed and CC/NCC systems in the {\tt \agn} case.}
\label{t:fit_YM}
\hspace*{0.3cm}{\footnotesize}
\begin{tabular}{lllllllll} 
\hline
   Simulation   & Sample   &  $A_{500}$ &    $B_{500}$  & $\sigma_{log_{10} Y_{500}}$  & $A_{2500}$ & $B_{2500}$ & $\sigma_{log_{10} Y_{2500}}$ \\
\hline 
{\tt \nr} 	& Complete &  $-4.282\pm0.008$ &   $1.688\pm0.012$ &    $0.062$ &   $-4.542\pm0.012$ &   $1.726\pm0.017$ &  $0.085$ \\ 
\\ 
{\tt \w} 	& Complete &   $-4.382\pm0.008$ &   $1.726\pm0.011$ &    $0.062$ &     $-4.712\pm0.013$ &   $1.851\pm0.019$ &  $0.098$\\ 
\\
{\tt \agn} 	& Complete &   $-4.305\pm0.009$ &   $1.685\pm0.013$ &    $0.067$ &     $-4.585\pm0.014$ &   $1.755\pm0.020$ &  $0.104$ \\ 
\\
		& Reduced-CC &   $-4.253\pm0.028$ &   $1.561\pm0.071$ &    $0.047$ &     $-4.584\pm0.052$ &   $1.747\pm0.144$ &  $0.081$ \\
		& Reduced-NCC &   $-4.291\pm0.015$ &   $1.618\pm0.037$ &    $0.055$ &     $-4.548\pm0.020$ &   $1.706\pm0.059$ &  $0.082$ \\
 		& Reduced-Regular &   $-4.316\pm0.013$ &   $1.639\pm0.025$ &    $0.032$ &   $-4.595\pm0.030$ &   $1.745\pm0.062$ &  $0.074$ \\
		& Reduced-Disturbed &   $-4.249\pm0.038$ &   $1.518\pm0.105$ &    $0.073$ &  $-4.485\pm0.031$ &   $1.681\pm0.107$ &  $0.076$ \\ 
\hline 
\end{tabular}

\label{tab:Y-Mfittings}
\end{table*}

In our simulations the core thermal properties or the cluster
dynamical state affect the scatter in the \ysz$-M$ relations.
Table~\ref{tab:Y-Mfittings} reports the best-fitting values obtained
for the power-law relation of Eq.~\ref{eq:power_law} when the relevant
quantities are computed within $R_{500}$ and $R_{2500}$. In
particular, we show results for the complete set of groups and
clusters in our three models and for the two distinct subsamples of
regular/disturbed and CC/NCC clusters in our {\tt \agn} simulations.

As discussed previously, when we consider the complete sample of
systems we obtain, for all models, a scatter of $6-7$ per cent at
$R_{500}$ and $8-10$ per cent at $R_{2500}$, with the {\tt \agn} model
showing the largest scatter. If we analyze instead the results
obtained for the subsamples of regular/disturbed and CC/NCC clusters
in our {\tt \agn} simulations we get interesting conclusions. At
$R_{500}$, we note that cluster dynamical state has a minor impact on
normalization and slope of the scaling relations. As for the scatter,
it is higher ($\simeq 7$ per cent) for the dynamically disturbed
clusters, whereas it clearly decreases for relaxed systems ($\simeq 3$
per cent). However, in the case of CC and NCC systems we see that the
scatter at $R_{500}$ is similar, with values of 
$\simeq 5$ and $5.5$ per cent, respectively. 

As for the results at $R_{2500}$ the {\tt \w} model 
has a higher slope and normalization with respect to the other two. At the scale 
of clusters with $10^{14} M_{\odot} < M_{2500} < 10^{15} M_{\odot}$, 
corresponding to the range of the observational samples of 
\citet{Bonamente_2008}, \citet{Czakon_15} and \citet{Marrone_2012}, 
the difference among models is reduced by a factor of $\sim 2.5$.
Considering only the disturbed subsample
gives a tilt to the relation, namely a slightly higher normalization than for
the regular systems. At the same time, the scatter obtained for
disturbed systems is close to that for relaxed clusters, with
values around $7-8$ per cent. In this
case, NCC clusters also show a similar scatter as CC
systems. 
This is consistent with the results on pressure profiles displayed in
Fig.~\ref{fig:profiles_cc_ncc} for CC and NCC clusters, showing a
similar scatter at radii below $0.1 R_{500}$.

\subsection{The evolution of the \ysz$-M$ relation}
\label{sec:sz_evol}

\begin{figure}
\includegraphics[width=9cm]{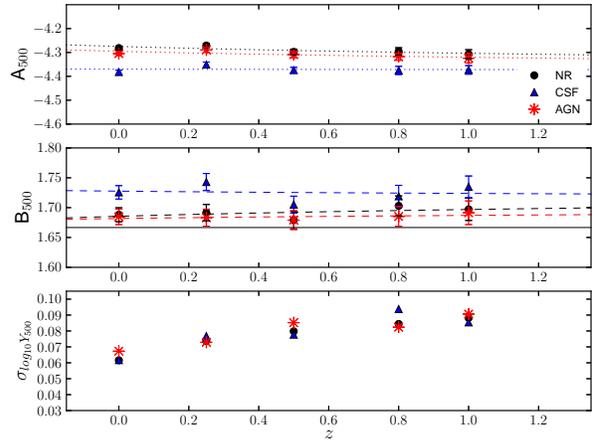}
\caption{Redshift evolution of the $Y_{\rm SZ,500}$--$M_{500}$ scaling
  relation for the complete set of groups and clusters in the
  {\tt \nr} (black circles), {\tt \agn} (red asterisks) and {\tt \w}
  (blue triangles) cases. In the three panels (from top down), we show
  the evolution of the normalization $A_{500}$, the slope,
  $B_{500}$, and the scatter, $\sigma_{log_{10} Y_{500}}$, of the relation.
  For the normalization and the slope we overplot, with dotted and
  dashed lines respectively, the best-fit relation for redshift
  evolution. The self-similar value for the slope is represented in
  the middle panel with a black continuous line.}
\label{fig:YMred}
\end{figure}

In Sect.~\ref{sec:zevol} we showed that pressure profiles for our
simulated massive clusters and observations from \citet{McDonald_2014}
do not show evidence of deviation from self--similar evolution from $z
= 0.5$ to $z=0.8$. In this section we will focus instead on the
redshift evolution of the best-fit parameters for the \ysz--$M$
relation up to $z=1$.  In Fig.~\ref{fig:YMred} we report the best-fit
values for $A_{500}$, $B_{500}$ and the scatter at different redshifts
for the {\tt \nr}, {\tt \w} and {\tt \agn} simulation sets.  To better
understand the evolution, we overplot our mean data for each redshift
with the best fit relation obtained for the normalization and the
slope of the relation, both normalized to the values at $z=0$, i.e.,
$A_0$ and $B_0$, respectively:
$$ A = A_{0} \; (1 + z)^{\alpha} $$
$$ B = B_{0} \; (1 + z)^{\beta} \, . $$

As shown in the top panel of Fig.~\ref{fig:YMred}, the normalization
$A_{500}$ remains almost constant with redshift. The radiative {\tt
  \w} simulation has the lowest normalization, in agreement with the
results reported in the left panel of Fig.~\ref{fig:Y-M500} at $z=0$.
As shown in previous works \citep[e.g.][]{Fabjan_2011,
Planelles_2013_b}, the residual variation of cluster properties in
radiative simulations with respect to {\tt \nr} ones is a consequence
of overcooling that removes gas from the hot phase. At the same time,
including AGN feedback partially prevents cooling and conversely
brings {\tt \nr} and {\tt \agn} normalizations to be closer.  The
middle panel of Fig.~\ref{fig:YMred} reports the slope $B_{500}$ for
the three models: while in the {\tt \nr} and {\tt \agn} case the slope is 
very close to the expected self-similar value ($B_{500} =5/3$), the {\tt \w}
simulation shows slightly steeper slopes at all redshifts with a small 
departure (less than $\sim 5\%$) from self-similarity that is constant 
with increasing redshift. 
Since the slope does not differ significantly with physics, the differences in 
normalization are effective and do not depend upon the choice of the pivotal point.
As for the scatter of the \ysz$-M$ relation, previous
findings from simulations \citep[e.g.][]{sembolini_etal13} show that the
scatter remains constant and at low levels. In our case 
the scatter has values around $0.05-0.09$ and increases with redshift, 
but if we focus only on massive clusters as in \cite{sembolini_etal13}, 
the scatter remains constant and at lower levels, $\sim 0.03-0.05$. 
A small change in the best fitting values for the
normalization and slope parameters as the simulation model is changed
was also noted by \citet{Kay_2012} for their set of simulated
clusters.

\subsection{The effect of clumpiness and the \ysz-\yx\ relation}
\label{sz_clumps}

In principle, X-ray measurements of gas mass and temperature can be
used to predict the SZ effect signal. In particular, providing that clusters
are isothermal, one can rewrite the integrated thermal pressure as
$Y_{SZ} D^2_A \propto T_e \int n_e dV = M_{gas} T_e$.  This argument
led \citet{Kravtsov_2006} to introduce the X--ray analogue of the SZ
Comptonization parameter, $Y_X$, defined as the product of the X--ray
derived gas mass and temperature. In principle, in observations, the comparison between
\ysz\ and \yx\ can provide information both on the inner ICM
thermal structure and on the clumpiness in gas density. In fact,
X-ray and SZ signals have a dependence on gas density, $n_e$, that
is, respectively, quadratic and linear and, therefore,
differences between the two signals 
arise in the presence of inhomogeneities in the gas distribution that would boost 
\yx\ with respect to \ysz\ \citep[e.g.][]{Giodini_2013}. Furthermore, any
inhomogeneity in the ICM thermal structure would induce a difference
between the electron temperature, that enters in \ysz, and the
spectroscopic temperature, which enters in \yx\
\citep[e.g.][]{Rasia_2014}. We remind here that to compute \ysz\ 
from simulations we use the mass-weighted temperature profile ($T_{mw}$), 
while to obtain \yx\ we employ the spectroscopic--like temperature
($T_{sl}$), as defined by \citet{Mazzotta_2004}, excising particles
within $0.15 R_{500}$.

\begin{figure}
\hspace{-0.8cm}
{\includegraphics[width=12.0cm]{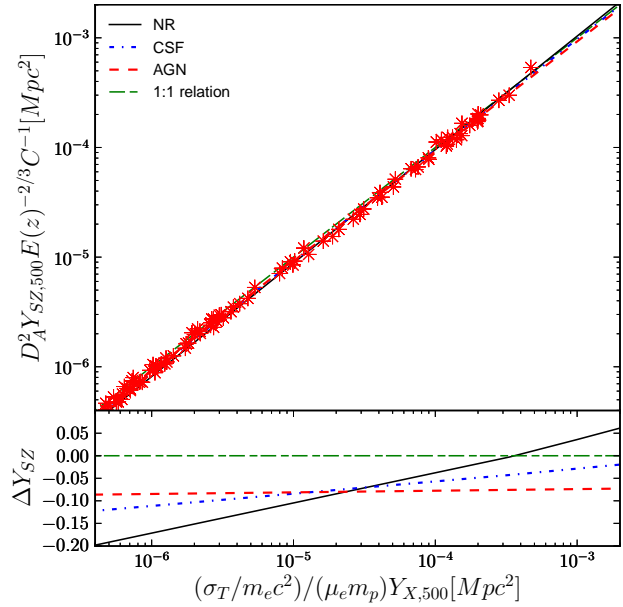}}
\caption{Upper panel: relation between the spherically integrated SZ flux and \yx\
  at $R_{500}$ for the complete sample of simulated systems. The SZ flux 
  is obtained by multiplying the cylindrical \ysz\ with the parameter C (described  
  in Section \ref{sz_clumps}).
  Best-fit relations are plotted with black continuous, blue dash-dotted and
  red dashed lines for the {\tt \nr}, {\tt \w} and {\tt \agn}
  simulations. For the sake of clarity, data (red asterisks) are plotted
  only for the {\tt \agn} case.  \yx\ is evaluated using
  core-excised ($0.15-1 R_{500}$) spectroscopic--like temperatures.
  The green long-short dashed line stands for the identity relation. 
  Lower panel: residuals of the \ysz-\yx\ relation computed 
  with respect to the identity relation, where 
  $\Delta Y_{\rm SZ} = (Y_{\rm SZ} - Y_{\rm SZ,identity}) / Y_{\rm SZ}$. 
  Same legend as in upper panel applies.
  }
\label{fig:Y-YX}
\end{figure}

Figure \ref{fig:Y-YX} shows the relation between the 
integrated thermal pressure \ysz\ and \yx, properly scaled to match
the same units. The relation is described by
\be 
Y_{SZ} D_A^2 = \frac{\sigma_T}{m_e c^2 m_p \mu_e} C Y_X \; ,
\ee
where $\mu_e=1.14$ is
the mean molecular weight per free electron  
and  $C$ is a factor used to properly account for the different domain 
of integration of \ysz\ and \yx. Here, the parameter 
$C$ was obtained for each cluster by dividing the cylindrical \ysz\ by
$Y_{\rm sph,500} = \frac{\sigma_T k_B}{m_e c^2} \int_0^{R_{500}} n_e(r) T_{mw}(r) dV$, 
the integral of the product between electron density and temperature profile 
inside a sphere of radius $R_{500}$. In all our models the parameter $C$ has 
a typical value of $\sim 1.4-1.5\pm0.2$ at $R_{500}$, in agreement with the value of 
$1.3\pm0.2$ reported by  \cite{sembolini_etal13}.

As before, we compute the best-fit relations following Eq.~\ref{eq:power_law}.
For our simulated
clusters we find that in the {\tt \nr} case clusters have the highest slope 
($B= 1.04, 1.01, 1.00$ for {\tt \nr}, {\tt \w} and {\tt \agn}, respectively), 
and both radiative models show the highest normalization 
($A=0.12, 0.03, -0.02$ for {\tt \nr}, {\tt \w} and {\tt \agn}, respectively).
Differences between the three models are of the order of at most $\sim 12\%$, 
as we can see in the bottom panel of the same figure, where we show the 
residuals of our models with respect to the theoretical identity relation.
In fact, we note that for our two radiative models the 
\ysz\ - \yx\ relation is close to the identity relation
(represented by the green long-short dashed line). In the case of the {\tt \agn} model 
we obtain a good agreement with the relation provided by \citet{Arnaud_2010}, that reports
a value of $Y_{sph}/CY_X = 0.924\pm0.004$ for the slope fixed to one.
Our simulated clusters show a scatter of $\sim 0.03-0.04$ in all models (for the sake of
clarity we only plot data for the {\tt \agn} case), being similar to what found
from observational data \citep[e.g.][]{Arnaud_2010}.

Since we use the true gas mass obtained from simulations in the
computation of both \yx\ and \ysz, their comparison provides
information on the deviation between $T_{sl}$, entering in the
computation of \yx, and $T_{mw}$, entering instead in \ysz\
\citep[e.g.][]{Biffi_2014}.  In this case, keeping in mind that the
innermost part of the cluster is excised when computing $T_{sl}$,
there is only a slight deviation from the identity relation that is
due to this temperature inhomogeneity.  Radiative simulations
generally show a smoother ICM thermal distribution than non-radiative
runs \citep[e.g.][]{Planelles_2013_b}.  In principle, as found by
\citet{Rasia_2014}, there should be a distinct behaviour in
non-radiative and radiative simulations that is a clear signature of
two counteracting processes: the cooling process, that removes the
low-entropy gas from the diffuse phase, and the heating by feedback
processes, that prevents the gas removal. When both effects are at
work, the temperature contrast between the clumps and the diffuse
medium is lower. At the same time, this effect should be particularly
relevant for clusters with larger masses, that are more affected by
inhomogeneities from gas accretion.  In our case, however, thanks to
the improved description of mixing of the new hydro scheme, both
radiative and non-radiative runs are quite close to each other, thus
indicating a more similar ICM thermal distribution in all cases.

As already mentioned, we compute  \yx\ using the true gas mass 
obtained from simulations, which however is not an observable.
For an estimate of the effect that the bias in clumping has 
on the \ysz-\yx\ relation we can use the data previously presented  in Section \ref{sec:clumping}.
In the two upper panels of Fig.~\ref{fig:clumping_z0} we have shown that the median 
gas density clumping factor, $C_{\rho}$, is of the order of $1-3$ within $R_{\rm 500}$ at the
scale of the groups and clusters of the reduced sample.
To obtain an estimate of how the \ysz-\yx\ relation would change 
due to the clumping, we compute the mean $C_{\rho}$ inside $R_{500}$ for clusters 
and groups separately. 
In this case, gas masses would increase by a factor 
$\sqrt{\langle C_{\rho}\rangle}$, which is equal to $1.051, 1.032, 1.032$ for groups 
($\langle M_{500}\rangle \sim 1.9 \times 10^{14} h^{-1}M_\odot$) and to $1.124, 1.076, 1.076$ 
for clusters ($\langle M_{500}\rangle \sim 7.7 \times 10^{14} h^{-1}M_\odot$), 
respectively in the {\tt \nr}, {\tt \w} and {\tt \agn} simulations.
For radiative simulations the bias at the group scale would translate into a $\sim3-4\%$ shift 
towards higher gas masses (or, equivalently, \yx), compatible with the scatter of our data, 
while the shift would be two times larger ($\sim 7.5\%$) at the level of clusters. 
For the non-radiative simulations the discrepancy would be, as expected, even larger ($\sim 5\%$
and $12\%$ for groups and clusters, respectively).   
Due to the different behaviour at the group and cluster scales,
the slope of the \ysz-\yx\ relation we would obtain would be shallower, with an estimated
value of $B \sim 0.85$ for all the three different physics.

\section{Conclusions}
\label{sec:summary}
We have performed a detailed analysis of the ICM thermal pressure
distribution of a sample of simulated galaxy clusters and groups, with
the main purpose of analysing the existing connection between pressure
profiles, ICM clumping, and SZ scaling relations.

Our sample of groups and clusters has been extracted from a set of
hydrodynamical simulations performed with a version of the
{\footnotesize {\sc GADGET-3}} code that includes an improved
description of SPH \citep{Beck_2015}. We analyzed three sets of
simulations: besides a non--radiative ({\tt \nr}) simulation set, we
also considered two sets of radiative simulations, a first one
including star formation and stellar feedback ({\tt \w} set), and a
second one also including the effect of gas accretion onto SMBHs and
the ensuing AGN feedback ({\tt \agn} set).  Based on different cluster
properties, we have classified the systems in our {\tt \agn}
simulations in CC and NCC halos \citep[see][for further
details]{Rasia_2015} and, depending on their global dynamical state,
in regular or disturbed systems \citep[see also][]{Biffi_2016}.  These
simulations, that we analyzed out to $z=1$, allow us to characterize
in detail the ICM pressure distribution, the ICM clumping, and the SZ scaling relations
as a function of redshift, ICM physics, cluster mass
and cluster dynamical state.

Our main findings can be summarized as follows:

\begin{itemize}
 
\item Independently of the physics included in our simulations, the
  mean pressure profiles obtained for our sample of groups and
  clusters show a good agreement with different
  observational samples \citep[e.g.][]{Arnaud_2010,
    Sun_2011,Planck_2013, Sayers_2013, McDonald_2014}.  
   In particular, in the case of clusters the agreement is 
   within 1-$\sigma$ at $r/R_{500}\mincir 1$, while in outer regions ($r \magcir R_{500}$)
    simulated profiles are systematically higher (by $\sim20$ per cent) than the universal profile presented in \citet{Arnaud_2010}.
    In the case of groups, our {\tt \agn} simulations produce pressure profiles that agree within $\sim 15-20\%$ 
with the observations reported by \citet{Sun_2011} and with the universal fit of \citet{Arnaud_2010}.
  This general consistency is also supported by a quantitative comparison between
  fitting parameters of the generalized NFW model.

\item When we analyze separately the pressure profiles of the samples
  of CC and NCC clusters in the {\tt \agn} set, we also obtain a
  good agreement with observed profiles of these two
  cluster populations \citep[e.g.][]{Sayers_2013,
   McDonald_2014}. Namely, whereas at intermediate and outer cluster
  radii pressure profiles of CC and NCC are similar (within a few per cent) to each other,
  they are clearly different in inner clusters regions,
  $r\le0.2R_{500}$, with CC clusters showing higher central pressure
  values (by a factor $\sim2.5$) and steeper profiles.  As shown by \citet{Rasia_2015}, this
  sample of CC and NCC systems also shows a good agreement with
  observational data in terms of entropy and iron abundance profiles.

\item In agreement with observational data
  \citep[e.g.][]{McDonald_2014, Adam_2014}, we obtain in all cases a
  redshift evolution of the pressure profiles of massive clusters that is
  consistent with the self-similar one, at least back to $z=1$.  The
  major discrepancies between simulated and observed profiles appear
  in very central and outer cluster regions, being in any case smaller
  than $\sim 10-15$ per cent.

\item In accordance with previous numerical studies
  \citep[e.g.][]{Roncarelli_2013, Battaglia_2014}, the gas density
  clumping derived in our simulations increases with the distance to
  the cluster center. In addition, its magnitude also increases with
  increasing redshift, with the mass of the considered systems and in
  dynamically unrelaxed objects.  A similar trend is also found for
  the gas pressure clumping, which shows, however, a lower magnitude.
 
\item While our {\tt \nr} simulations produce quite higher values of the
  gas density clumping ($\magcir 20\%$ beyond $R_{200}$), 
  both of our radiative runs produce lower values ($\sqrt C_{\rho}\sim 1.2$ at $R_{200}$). 
  While this level of clumping is in good match with, for instance,
  the observational estimate by \citet{Eckert_2013_2}, it is
  significantly smaller than that inferred to explain reported
  measurements of high entropy in the outskirts of clusters
  \citep[e.g.][]{Simionescu2011}.

\item The $Y_{SZ}-M$ scaling relation of our simulations is in good
  agreement with observational data at the scale of massive clusters
  \citep[e.g.][]{Bonamente_2008, Marrone_2012, Czakon_15}.  In addition,
  consistently with previous numerical analyses
  \citep[e.g.][]{Kay_2012, sembolini_etal13, Pike_2014}, our results
  at $R_{500}$ do not show any significant dependence on the physics
  included.
  Moreover, for all our models, normalization and slope of the relation 
  do not show any clear redshift evolution, but only an increase in scatter 
  (from $6$ to $9\%$ for $z=0$ to $1$).
  
\item As for the scatter in the $Y_{SZ}-M$ relations, we predict a
  clear dependence on the cluster dynamical state, with disturbed
  clusters showing $\sim 2$ times larger scatter than regular systems at $R_{500}$. 
  This finding is not reflected in the CC/NCC samples confirming that 
  the cluster dynamical state, measured at larger scales, is not strongly 
  correlated with its cool-coreness \citep[][]{Meneghetti_2014, Donahue_2016}.

\end{itemize}

While observations and simulations agree in that galaxy clusters
behave as a homogeneous population of objects at intermediate radii,
$0.15R_{500}<r<R_{500}$,  cluster core regions and cluster outskirts 
are quite sensitive to a number of dynamical and
feedback processes. Given their definitions, X-ray and SZ
observations provide a complementary description of the ICM
thermodynamics, with the former being better suited to prove inner
high-density cluster regions, while the latter being more sensitive to
outer low-density regions.  From an observational point of view,
however, constraining the outskirts of clusters is challenging because
they require observations with high sensitivity: long exposures to
detect low surface brightness and sharp point spread functions 
(PSF) to remove contributions from point-like sources 
\citep[e.g.][for a recent review]{Reiprich_2013}.
In particular, although ICM clumping can affect derived X-rays and SZ
cluster properties, current estimations are still uncertain.  A
detailed analysis of cluster outskirts will require the next
generation of high-sensitivity X-ray observatories, such as
Athena\footnote{\tt http://www.the-athena-x-ray-observatory.eu}
\citep[e.g.][]{Pointecouteau_2013, Nandra_2013, Ettori_2013_b}, in
combination with high-resolution SZ observations.  Simulations, like
those presented in this paper, will represent an ideal interpretative
framework to unveil the physical processes determining the structure
of the ICM pressure, also in view of an improved calibration of galaxy
clusters as precise tools for cosmology.

\section*{ACKNOWLEDGEMENTS} 
 
The authors would like to thank Volker Springel for allowing us to
access the developer version of the {\small GADGET} code, and Matteo
Costanzi for helping us in the use of the MCMC algorithm.  
We also would like to thank Mauro Roncarelli, Daisuke Nagai, Dominique Eckert, 
Jack Sayers, and Michael McDonald for useful comments and discussions as well as the 
anonymous referee for constructive comments that helped to improve the 
presentation of our results.

We acknowledge financial support from PIIF-GA- 2013-627474, NSF AST-1210973, 
PRIN-MIUR 201278X4FL, PRIN-INAF 2012 ``The Universe in a Box: Multi-scale 
Simulations of Cosmic Structures'', the INFN INDARK grant, ``Consorzio per la Fisica'' of Trieste, 
CONICET-Argentina, FonCyT. 
SP also acknowledges support by the {\it Spanish Ministerio de Econom{\'i}a y 
Competitividad} (MINECO, grant AYA2013-48226-C3-2-P) and the Generalitat Valenciana (grant GVACOMP2015-227).
DF acknowledges partial support by the {\it Slovenian Research Agency}.
Simulations are carried out using Flux HCP Cluster at the University of Michigan, Galileo at 
CINECA (Italy), with CPU time assigned through ISCRA proposals and an agreement with 
the University of Trieste, and PICO at CINECA through our ISCRA project IscrC\_{GALPP2}.
CRF acknowledges founding from CONICET, FonCyT and SeCyT-UNC, Argentina.
AMB is supported by the DFG cluster of excellence ``Universe'' and by the DFG Research Unit 
1254 ``Magnetization of interstellar and intergalactic media'' and by the Leibniz-Rechenzentrum 
via project ``pr92ju''.
MG is supported by NASA through Einstein Postdoctoral Fellowship Award Number PF-160137 
issued by the Chandra X-ray Observatory Center, which is operated by the SAO for and on behalf of 
NASA under contract NAS8-03060.

\bibliographystyle{mnbst}
\bibliography{PressurePaper}

\end{document}